%% file: ms.tex
\documentclass{emulateapj}

\usepackage{lscape}

\newcommand{\eps}[1]{\mbox{log~$\epsilon$\,(#1)}}

\newcommand\wave[1]{\mbox{#1\,\AA}}
\def\bdcowan{\mbox{BD$+$17~3248}}
\def\cssneden{\mbox{CS~22892--052}}
\def\cshayek{\mbox{CS~29491--069}}
\def\cshonda{\mbox{CS~30306--132}}
\def\cslai{\mbox{CS~31078--018}}
\def\cshill{\mbox{CS~31082--001}}
\def\hehayek{\mbox{HE~1219--0312}}
\def\hefrebel{\mbox{HE~1523--0901}}

\def\kmsec{\mbox{km~s$^{\rm -1}$}}

\def\rpro{\mbox{$r$-process}}
\def\spro{\mbox{$s$-process}}
\def\ncap{\mbox{$n$-capture}}
\def\second{{2$^{\rm nd}$}}
\def\third{{3$^{\rm rd}$}}

\shorttitle{Lead and Thorium from the $r$-process in Metal-Poor Stars}
\shortauthors{Roederer et al.}
\submitted{Accepted for publication in the Astrophysical Journal}

\begin{document}

\title{The End of Nucleosynthesis:
Production of Lead and Thorium in the Early Galaxy}

\author{
Ian U.\ Roederer\altaffilmark{1},
Karl-Ludwig Kratz\altaffilmark{2},
Anna Frebel\altaffilmark{3}, 
Norbert Christlieb\altaffilmark{4},
Bernd Pfeiffer\altaffilmark{2},
John J.\ Cowan\altaffilmark{5},
Christopher Sneden\altaffilmark{1} 
}

\altaffiltext{1}{Department of Astronomy, University of Texas at Austin,
1 University Station, C1400, Austin, TX 78712-0259; 
iur,chris@astro.as.utexas.edu}

\altaffiltext{2}{Max-Planck-Institut f\"{u}r Chemie, Otto-Hahn-Institut,
J.-J.-Becherweg 27, D-55128 Mainz, Germany; klk,bpfeiffe@uni-mainz.de}

\altaffiltext{3}{Harvard-Smithsonian Center for Astrophysics,
60 Garden St., MS-20, Cambridge, MA 02138; afrebel@cfa.harvard.edu}

\altaffiltext{4}{Zentrum f\"{u}r Astronomie der Universit\"{a}t
Heidelberg, Landessternwarte, K\"{o}nigstuhl 12, D-69117 Heidelberg, 
Germany; N.Christlieb@lsw.uni-heidelberg.de}

\altaffiltext{5}{Homer L.\ Dodge Department of Physics and Astronomy, 
University of Oklahoma, Norman, OK 73019; cowan@nhn.ou.edu}

\begin{abstract}

We examine the Pb and Th abundances 
in 27 metal-poor stars ($-3.1 <$~[Fe/H]~$<-1.4$)
whose very heavy metal ($Z > 56$) enrichment was produced 
only by the rapid ($r$-) nucleosynthesis process.
New abundances are derived from HST/STIS, Keck/HIRES, and VLT/UVES spectra
and combined with other measurements from the literature
to form a more complete picture of nucleosynthesis 
of the heaviest elements produced in the $r$-process.
In all cases, the abundance ratios among the rare earth elements
and the \third\ $r$-process peak elements considered (La, Eu, Er, Hf, and Ir)
are constant and equivalent to the scaled solar system $r$-process
abundance distribution.
We compare the stellar observations with $r$-process calculations within the
classical ``waiting-point'' approximation. 
In these computations a superposition of 15 weighted neutron-density 
components in the range 23~$\le$~log~n$_n$~$\le$~30 
is fit to the $r$-process abundance peaks to successfully 
reproduce both the stable solar system isotopic distribution 
and the stable heavy element abundance pattern 
between Ba and U in low-metallicity stars. 
Under these astrophysical conditions, which are typical
of the ``main'' $r$-process, we find very good agreement between the 
stellar Pb \rpro\ abundances and those predicted by our model.
For stars with anomalously high Th/Eu ratios (the so-called actinide boost), 
our observations demonstrate that 
any nucleosynthetic deviations from the main \rpro\ 
affect---at most---only the elements beyond the \third\ $r$-process peak, 
namely Pb, Th, and U.
Our theoretical calculations also indicate that possible $r$-process
abundance ``losses''
by nuclear fission are negligible for isotopes along the $r$-process 
path between Pb and the long-lived radioactive isotopes of Th and U.

\end{abstract}

\keywords{
Galaxy: halo ---
nuclear reactions, nucleosynthesis, abundances ---
stars: abundances ---
stars: Population II
}

\section{Introduction}
\label{introduction}

Nucleosynthesis of the heaviest elements in the Universe is 
accomplished by successive additions of neutrons to existing 
iron (Fe) group nuclei in stars.
Two factors restrict a star's ability to produce very heavy nuclei
through standard fusion reactions: the endothermic nature of fusion
reactions for species heavier than $^{56}$Fe and the
increased Coulomb barriers that discourage charged particle reactions
in isotopes with sufficiently high $Z$.
Isotopes heavier than those of the Fe group are therefore overwhelmingly
produced by neutron- ($n$-) capture processes.
Either a single neutron can be added on timescales longer than
the mean time to $\beta^{-}$ decay (the slow [$s$] \ncap\ reaction)
or many neutrons will be added 
before multiple $\alpha$ and $\beta^{-}$ decays return the isotope to 
stability (the rapid [$r$] \ncap\ reaction).
The rate of neutron captures and the resulting abundance patterns 
are strongly regulated by the 
physical conditions and neutron densities at the time of nucleosynthesis.

Understanding the production of the heaviest elements in the Universe
provides crucial insight into the nature of these processes.
An exact site for the location of the \rpro\ has not been conclusively
identified, although the appearance of \rpro\ material in stars of
very low metallicity ([Fe/H]~$= -3.0$)\footnote{
We adopt the standard spectroscopic notations that
[A/B]~$\equiv$ log$_{10}$(N$_{\rm A}$/N$_{\rm B}$)$_{\star}$~--
log$_{10}$(N$_{\rm A}$/N$_{\rm B}$)$_{\odot}$ and
log~$\epsilon$(A)~$\equiv$ 
log$_{10}$(N$_{\rm A}$/N$_{\rm H}$)~$+$~12.00
for elements A and B.}
argues against possible astrophysical sites (e.g., neutron-star 
mergers) that require long evolutionary timescales 
\citep[e.g.,][]{argast04}, and  
implies instead that some association with Type~II core collapse supernovae
(SNe) is likely \citep[e.g.,][]{cowan04,cowan06,farouqi09}.
The lack of a precise identification of the \rpro\ site, however, has
complicated efforts to model it, necessitating vast amounts of input nuclear
data and increasingly more sophisticated model approaches 
\citep[e.g.,][]{cowan91a,kratz93,chen95,freiburg99,pfeiffer01,kratz07b,arnould07}.
Nevertheless, encouraging progress has been made in recent years to 
experimentally determine the half-lives, nuclear masses, and \ncap\
cross-sections for nuclei along the \rpro\ path 
\citep{kratz93,rauscher00,pfeiffer01,kratz07b,schatz08},
and some astrophysically-motivated models of the \rpro\ have 
been able to reproduce the robust \rpro\ pattern between 
$A \simeq$~120 and the actinide region \citep[e.g.,][]{kratz07a,farouqi09}.
Low- and intermediate-mass stars ($\sim 1.5$--3.0$M_{\odot}$) that pass
through the asymptotic giant branch phase of evolution are the primary
source of \spro\ material \citep[e.g.,][]{busso99,lattanzio05,straniero06},
and investigators have had great success in matching model predictions
to observed \spro\ abundance patterns 
(e.g., \citealt{sneden08} and references therein).
Understanding these processes then
permits the study of additional astrophysics, such as
the use of radioactive isotopes to determine the ages of individual
stars and stellar systems \citep[e.g.,][]{cowan91,truran02}.
 
Lead (Pb) and bismuth (Bi) are the two heaviest nucleosynthesis products
of the \spro.
Pb has four naturally-occurring isotopes:
$^{204}$Pb, $^{206}$Pb, $^{207}$Pb, and $^{208}$Pb;
Bi has one, $^{209}$Bi.
The $^{204}$Pb isotope is blocked from \rpro\ production by the
stable nucleus $^{204}$Hg and will not be considered further here.
$^{208}$Pb is a double-magic nucleus ($N=$~126 and $Z=$~82,
closing both its neutron and proton shells), 
which significantly lowers its cross section to further \ncap.
One subsequent neutron capture (and $\beta^-$ decay)
produces the stable atom $^{209}$Bi
and another produces $^{210}$Bi, which may either
(in its isomeric state) $\beta^-$ decay or 
(in its unstable but relatively long-lived ground state, 
$t_{\rm 1/2}= 3\times 10^{6}$~yr) 
$\alpha$ decay.
Both processes produce $^{206}$Pb.
Thus, a Pb-Bi terminating cycle forms \citep{burbidge57,clayton67,ratzel04}.

In low metallicity stars the ratio of
free neutrons to Fe-peak seeds is higher,
and on average more neutrons are captured per seed nucleus
\citep{malaney86,clayton88}.
This results in a high fraction of heavy \spro\ nuclei being produced 
\citep{gallino98,busso99,busso01,goriely01,travaglio01,vaneck01,cristallo09}.
Alternatively, an extended period of \spro\ nucleosynthesis
in a given star could also produce a large number of heavy nuclei.
These properties of \spro\ nucleosynthesis have led to
spectacular Pb enhancements in several metal-poor stars, such as
\mbox{CS~29526--110} ([Pb/Fe]~$=+3.3$, \citealt{aoki02}),
\mbox{HD~187861} ([Pb/Fe]~$=+3.3$, \citealt{vaneck03}), 
\mbox{HE~0024--2523} ([Pb/Fe]~$+3.3$, \citealt{lucatello03}), 
\mbox{SDSS~0126+06} ([Pb/Fe]~$+3.4$, \citealt{aoki08a}), and
\mbox{CS~29497-030} ([Pb/Fe]~$=+3.65$, \citealt{ivans05}).

The termination of the \rpro\ occurs very differently.
Nuclei along the \rpro\ path 
with $A \gtrsim 250$ will undergo (spontaneous, neutron-induced 
and $\beta$-delayed) fission
and repopulate the \rpro\ chain at lower nuclear masses
(Panov et al. 2008).
Neutron-rich nuclei produced in the \rpro\ that have closed neutron shells at 
$N =$~ 126 (at the time of the termination of the \rpro)
will rapidly $\beta^-$ decay to atoms of elements
at the \third\ \rpro\ peak---osmium (Os), iridium (Ir), and 
platinum (Pt), but not Pb.
The nuclei that form with $A =$~206--208 nucleons in the \rpro\ 
(i.e., those that will $\beta^{-}$ decay to the Pb isotopes)
have neither a magic number of neutrons or protons, 
so they are produced in smaller relative amounts.
Even in the metal-poor stars with the most extreme \rpro\ overabundances,
the [Pb/Fe] ratios are extremely depressed.
Pb produced exclusively in the \rpro\ 
has been detected previously in only three stars with [Fe/H]~$<-2.0$
(\mbox{HD~214925}, [Fe/H]~$= -2.0$, \citealt{aoki08b};
\mbox{HD~221170}, [Fe/H]~$= -2.2$, \citealt{ivans06};
\cshill, [Fe/H]~$= -2.9$, \citealt{plez04}).
This overall lack of observational data concerning Pb abundances in 
very metal-poor, 
\rpro\ enriched stars serves as one motive for our present study.

There is an additional nucleosynthesis path for three of the Pb isotopes.
All nuclei heavier than $^{209}$Bi are ultimately unstable to 
$\alpha$ or $\beta$ decay and will follow decay chains that 
leave them as either Pb or Bi.
In this way, the Pb and Bi abundances in a star enriched by \rpro\ material
will increase with time as these heavier nuclei gradually decay.
Only two isotopes of thorium (Th) and uranium (U), 
$^{232}$Th and $^{238}$U, have halflives longer than 1~Gyr
($t_{\rm 1/2} [^{232}{\rm Th}] = 14.05\,\pm\,$0.06~Gyr and 
$t_{\rm 1/2} [^{238}{\rm U}]= 4.468\,\pm\,$0.003~Gyr;
\citealt{audi03} and references therein).
In old, metal-poor stars, these are the only heavier atoms
that we have any hope of observing today.
Only one detection of Bi has ever been made in a metal-poor
star, in the strongly \spro-enriched star \mbox{CS~29497-030} 
\citep{ivans05}.
This transition of Bi~\textsc{i} (\wave{3067}) 
lies very near the atmospheric transmission cutoff in a very crowded 
spectral region.
Similarly, U is very difficult to detect in metal-poor stars owing to its
relatively small abundance and the fact that the one useful
transition of U~\textsc{ii} in the visible spectrum (\wave{3859})
is severely blended with CN molecular features and lies on the wing
of a very strong Fe~\textsc{i} line.
Measurements of the U abundance have only been reported for 
three metal-poor stars.

Rather than focus on Bi or U, we investigate Pb and Th
to better understand the heaviest long-lived products 
of \rpro\ nucleosynthesis.
In this study we examine the Pb and Th abundances in a sample of
47 stars, deriving new abundances for 14 stars and adopting
literature values for the others.
We compare our results with \rpro\ nucleosynthesis model predictions
and discuss their implications for 
stellar ages and the chemical enrichment of the Galaxy.

\section{Observational Data}
\label{data}

For 12 stars in our sample,
we employ the same spectra used by \citet{cowan05}.
These data were collected with the Space Telescope Imaging Spectrograph
(STIS) on the \textit{Hubble Space Telescope} (\textit{HST}) and the
High Resolution Echelle Spectrograph (HIRES; \citealt{vogt94})
on the Keck~I Telescope.
Most, but not all, of these 12 stars are covered in both sets 
of observations; Table~\ref{atmtab} indicates which stars were observed 
with these two facilities.
The STIS spectra cover the wavelength range 
2410~$< \lambda <$~3070\,\AA\ at a resolving power 
$R \equiv \lambda/\Delta\lambda \sim$~30,000 and 
S/N~$\gtrsim$~50/1.
The HIRES spectra cover the wavelength range 
3160~$< \lambda <$~4600\,\AA\ at a resolving power
$R \sim$~45,000 and 
30/1~$\lesssim$~S/N~$\lesssim$~200/1.

We also derive abundances from two stars observed with UVES on the VLT, 
analyzed previously by \citet{hayek09}.
These spectra cover 3300~$< \lambda <$~4500\,\AA\
at resolving powers of $R \sim$~57,000 (\cshayek) and $R \sim$~71,000
(\hehayek).

We supplement our sample with measurements from the literature.
We include 5 stars from \citet{johnson01} and \citet{johnson02a},
11 stars from \citet{aoki08b}, and one star each from
\citet{hill02} and \citet{plez04}, 
\citet{honda04},
\citet{christlieb04},
\citet{ivans06}, 
\citet{frebel07}, and
\citet{lai08}.
We also include recent Pb or Th abundances for the globular clusters
M5 (2 stars; \citealt{yong08a,yong08b}),
M13 (4 stars; \citealt{yong06}), 
M15 (3 stars; \citealt{sneden00}), and
M92 (1 star; \citealt{johnson01} and \citealt{johnson02a}). 
Th has also been measured in one star in one dwarf spheroidal (dSph)
galaxy, Ursa Minor (UMi; \citealt{aoki07}).
Including these additional stars increases our sample to 47 stars.

\section{Abundance Analysis}
\label{analysis}

For the 14~stars whose abundances are reexamined here,
we adopt the model atmospheres derived 
by \citet{cowan02}, \citet{sneden03}, \citet{simmerer04}, 
\citet{cowan05}, and \citet{hayek09}.
These parameters are listed in Table~\ref{atmtab}.
We perform our abundance analysis using the most recent version of the
1D LTE spectral analysis code MOOG \citep{sneden73}.
We compute model atmospheres without convective overshooting 
from the \citet{kurucz93} grid, using
interpolation software developed by A.~McWilliam and I.~Ivans
(2003, private communication).

In addition to Pb and Th,
we also derive abundances of lanthanum (La), europium (Eu), 
erbium (Er), hafnium (Hf), and Ir.
Our goal is to sample the \ncap\ abundance ratios at regular
intervals in $N$ (or $Z$) using species that can be relatively
easily and reliably derived from ground-based blue and near-UV spectra.
Abundances derived from individual transitions are listed in 
Tables~\ref{indivtab1} and \ref{indivtab2}, along with the
relevant atomic data.
Some studies predate the most recent laboratory measurements of 
\ncap\ transition probabilities, so we update the abundances 
from those studies.
We adopt the latest $\log(gf)$ values for the \ncap\ species
from a number of recent studies:
La \citep{lawler01a},
Eu \citep{lawler01b},
Er \citep{lawler08},
Hf \citep{lawler07},
Ir \citep{ivarsson03},
Pb \citep{biemont00}, and
Th \citep{nilsson02}.
Final abundances for all elements, including updates of the literature
values, are listed in Table~\ref{abundtab}.

\subsection{Lead}
\label{pbanalysis}

We examine three Pb~\textsc{i} lines in each of our 14~stars: 
2833, 3683, and \wave{4057}.
The \wave{2833} resonance line is the strongest transition in 
stellar spectra, which demands use of the STIS UV spectra;
all other transitions are accessible from spectra obtained with
ground-based facilities.

Only one of the naturally-occurring Pb isotopes, 
$^{207}$Pb, exhibits hyperfine structure due to its
non-zero nuclear spin.
\citet{simons89} measured the isotope energy shifts and relative 
hyperfine splittings from FTS spectra.
We adopt their measurements and the Solar System (S.S.) isotopic
fractions \citep{lodders03} in our syntheses.
The \spro\ produced the majority of the Pb in the S.S.
In our stellar sample, where the \rpro\ is the dominant contributor
to the \ncap\ material, one might expect that a very different 
isotopic mix could affect our derived Pb abundances.
We find, however, that no sensible variations in the 
isotopic fractions can be detected in any of our Pb lines
due to the relatively small isotope shifts ($\lesssim$~0.02--0.03~\AA)
and overall weakness of the lines.

The \wave{2833.053} Pb~\textsc{i} 
line is blended with an Fe~\textsc{ii} line at 
\wave{2833.086}.
A $\log(gf)$ value for this line is listed in the critical compilation
of \citet{fuhr06}, who quote a ``C'' accuracy in $\log(gf)$;
i.e., $\pm$25\% or about 0.1~dex.
Another unidentified line at \wave{2832.91} mildly blends with the 
blue wing of the Pb line, but this has little effect on the derived
Pb abundance.
Both of these blending features are weak in stars with [Fe/H]~$<-$2.0
and are negligible in the most metal-poor stars of our sample.
The \wave{3683.464} Pb~\textsc{i} 
line sits on the wing of a Balmer series transition.
We empirically adjust the strength of this Balmer line to reproduce 
the continuous opacity at the location of the Pb line.  
An Fe~\textsc{i} line at \wave{3683.61} only slightly blends 
the red wing of the Pb line; we derive a $\log(gf)$ value of $-$1.95
for this line from an inverted solar analysis.
The \wave{4057.807} Pb~\textsc{i} 
line is heavily blended with CH features.
In typical \rpro\ enriched stars (with solar or subsolar C/Fe and 
approximately solar Pb/Fe), the Pb may be marginally detectable; 
if a star has supersolar C it is nearly impossible to identify 
absorption from Pb.

\begin{figure}
\epsscale{1.15}
\plotone{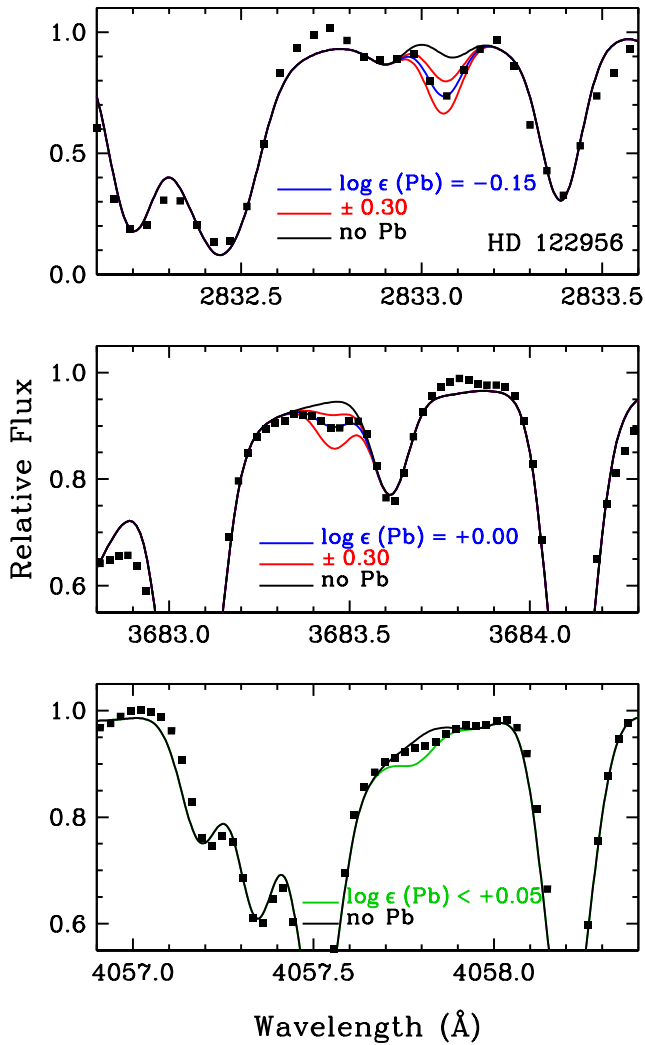} 
\caption{
\label{pbspec}
Syntheses of the three Pb~\textsc{i} lines in \mbox{HD~122956}.
The observed spectrum is indicated by black squares.
In the top two panels, 
our best-fit synthesis is indicated by the blue line.
Changes to this synthesis of $\pm$\,0.30~dex are indicated in red.
A synthesis with no Pb present is indicated by the black line.
We only derive an upper limit for Pb from the \wave{4057} line,
indicated by the green line in the bottom panel.
}
\end{figure}

We measure the Pb abundance in five stars in our sample, all
from either the 2833 or \wave{3683} lines.  
For all non-detections we report upper limits on the Pb abundance.
These individual measurements are listed in Tables~\ref{indivtab1}
and \ref{indivtab2}.
In Figure~\ref{pbspec} we display these three Pb lines in 
\mbox{HD~122956},
where we derive an abundance from the 2833 and \wave{3683} lines
and an upper limit from the \wave{4057} line.
Our choice of continuum normalization most affects the abundance
of the \wave{2833} line due to the relative lack of line-free 
continuum regions nearby.
Our stated uncertainties account for uncertainties in the 
continuum placement.

\subsection{Thorium}
\label{thanalysis}

We examine four Th~\textsc{ii} lines in each of our 14~stars:
3539, 4019, 4086, and \wave{4094}.
All of these transitions arise from the ground state.
The \wave{4019.129} Th~\textsc{ii} line is relatively strong but suffers 
from a number of blends.
Extensive reviews of the blending features have been made previously
(e.g., \citealt{lawler90,morell92,sneden96,norris97,johnson01}).
We provide only minor updates to these analyses.
An Fe~\textsc{i} line sits at \wave{4019.04}; \citet{fuhr06} report
a $\log(gf)$ for this line, scaled from \citet{may74}.  
They report an ``E'' accuracy for this line; 
i.e., $\pm>50$\% or $\gtrsim$~0.2--0.3~dex.
We allow the strength of this line to vary within these amounts to 
match the observed spectrum just blueward of the Th line.
\citet{lawler90} determined the $\log(gf)$ for a Co~\textsc{i} 
line at \wave{4019.126}, though this line was not found to 
contribute significantly in the very metal-poor stars examined here.
The strength of the hyperfine split
Co~\textsc{i} line at \wave{4019.3} can be treated as a free parameter 
to match the observed spectrum just redward of the Th line.
\citet{sneden96} identified a Ce~\textsc{ii} blend at \wave{4019.06}
in \cssneden\ that may explain extra absorption in the blue wing
of the Th line. 
We adopt an empirical $\log(gf)=-0.10$ for this line, which matches the 
observed spectrum in \bdcowan.
The Nd~\textsc{ii} line at \wave{4018.82}, for which 
\citet{denhartog03} report an experimental $\log(gf)$ value, can be 
used to estimate the Nd abundance.
We then set the Ce abundance from the typical Nd/Ce ratio in 
\rpro\ enriched metal poor stars, [Nd/Ce]~$\approx +$0.25.
In stars with an extreme overabundance of \ncap\ material, such as \cssneden, 
\citet{sneden96} note that exclusion of the Ce blend would only
increase the Th abundance by $\approx$~0.05~dex.
This would almost certainly decrease in stars with less severe
\ncap\ overabundances.

\begin{figure}
\epsscale{1.15}
\plotone{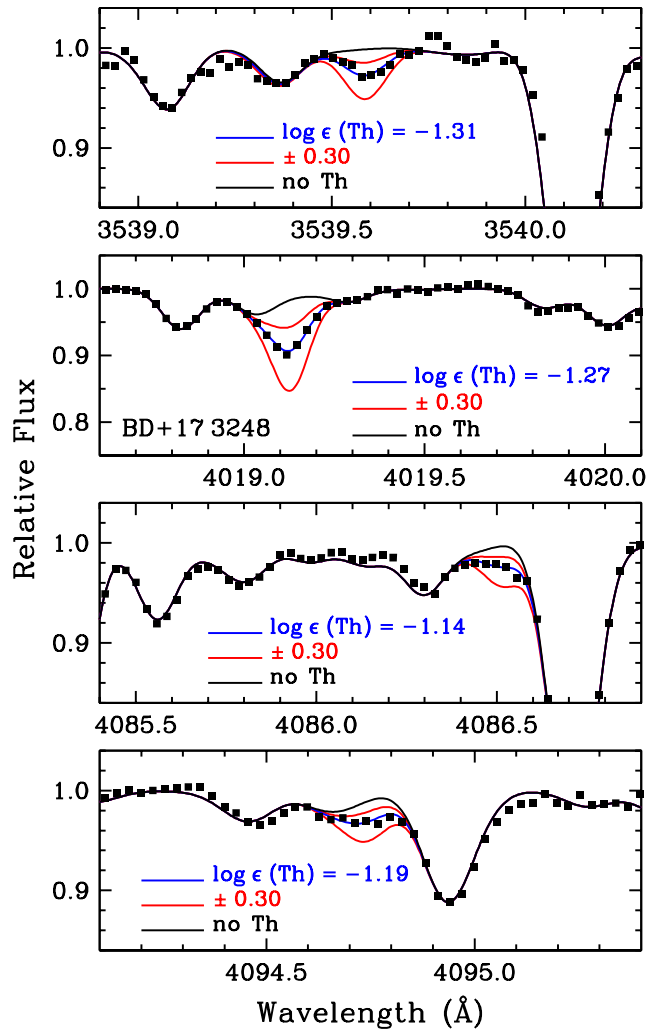} 
\caption{
\label{thspec}
Syntheses of the four Th~\textsc{ii} lines in \bdcowan.
Symbols are the same as in Figure~\ref{pbspec}.
We derive a Th abundance from each of these four lines in this star.
}
\end{figure}

The final serious contaminant to the \wave{4019} Th line is the 
$B^2\Sigma^- -X^2\Pi$(0--0) $^{13}$CH transition doublet 
(P$_{11}[16]$ and P$_{22}[16]$; \citealt{kepa96}), 
which was first discussed by \citet{norris97}.
It is thus necessary to estimate the C abundance 
and the $^{12}$C/$^{13}$C ratio in each of our stars.
The corresponding $^{12}$CH doublet 
lies approximately \wave{1} redward of the $^{13}$CH doublet.
\citet{norris97} have noted that the
wavelengths of these features may be incorrect in the Kurucz lists
by $\sim$0.15--0.25\AA.
We determine the absolute position of each of these lines
from the $^{12}$CH absorption lines, 
but we do not change the isotope shift 
between the $^{12}$CH and $^{13}$CH features.
We empirically set the overall strength from the $^{12}$CH doublet, 
adjusting the $\log(gf)$ values for the $^{12}$CH transitions and the
$^{13}$CH transitions together. 
We then employ these same steps with the $^{13}$CH and $^{12}$CH
doublets at $\approx$~4006 and \wave{4007}, respectively
(P$_{11}[15]$ and P$_{22}[15]$).
The $^{13}$CH doublet is relatively unblended here, permitting
measurement of the $^{12}$C/$^{13}$C ratio.
The overall strength of the contaminating \wave{4019} $^{13}$CH
feature can then be reduced by the $^{12}$C/$^{13}$C ratio.
Results derived from this method agree well with results derived 
from the CH linelist of B.\ Plez (2007, private communication).
Our measurements of $^{12}$C/$^{13}$C are listed in Table~\ref{cratiotab}.

The \wave{3539.587} Th~\textsc{ii} line is relatively weak but unblended.
The red wing of the \wave{4086.521} Th~\textsc{ii} line marginally
blends with the blue wing of a strong La~\textsc{ii} feature at 
\wave{4086.71}.
The blue wing of the \wave{4094.747} Th~\textsc{ii} line blends with
several features, including a Nd~\textsc{ii} line at 
\wave{4094.63}, an Er~\textsc{ii} line at \wave{4094.67}, and a 
$^{12}$CH line at \wave{4094.70}.  
Reliable $\log(gf)$ values are not known for any of these features.
The strength of these blends can be adjusted empirically in most stars; 
in other cases we instead measure an upper limit for the Th abundance.
A Ca~\textsc{i} line at \wave{4094.93} also mildly blends the red wing
of the Th line, but this line can be easily accounted for in our syntheses.
Syntheses for these four lines are shown in Figure~\ref{thspec}.

\subsection{Uncertainties and Comparisons to Previous Studies}
\label{uncertainties}

Pb is one of several \ncap\ species observed in the optical
regime in the neutral state in metal-poor stellar atmospheres, 
due to its relatively
high first ionization potential (I.P.), 7.42~eV.
The first I.P.\ of atoms of the \third\ \rpro\ peak (e.g., Os, Ir, Pt) 
is even higher (8.35, 8.97, and 8.96~eV, respectively), 
and these species, too, are 
observed in the neutral state in metal-poor stars.
Th, with a lower I.P.\ of 6.31~eV, is observed in its singly-ionized state,
as are all of the rare earth species.
The singly-ionized states of these atoms are the dominant species 
in typical metal-poor stellar atmospheres---even for Pb and the
\third\ \rpro\ peak elements---but less so than Th or the rare earth elements.
Thus the abundances derived for Th and the rare earth elements 
are determined from the majority species, but we caution that this
is not the case for Pb, where most of the atoms are not in the neutral state.
Further exploration of this issue is beyond the scope of the present work,
but we would welcome more detailed atomic model calculations of the Pb
ionization balance.

Species in different ionization states clearly respond differently
to conditions in the stellar atmosphere.
\citet{cowan05} present an extended discussion of the
uncertainties between ratios of elements with differing ionization states.
To summarize those results, for atmospheric uncertainties of 
$\Delta T_{\rm eff} = \pm 150$~K, 
$\Delta \log g = \pm 0.3$~dex, and 
$\Delta v_t = \pm 0.2$~\kmsec,
the total uncertainties in 
ratios between neutral and singly-ionized species (e.g., Pb/Eu or Th/Pb)
are typically $\simeq \pm 0.20$~dex.
Uncertainties in ratios between species of the same ionization state 
(e.g., Th/Eu) are much smaller, typically $\lesssim \pm 0.05$~dex.
When considering ratios between species of different ionization states,
we add an additional 0.20~dex uncertainty in quadrature with the 
individual measurement uncertainties.

Uncertainties in the abundance ratios resulting from neglect of the
electron scattering contribution to the continuous opacity should be
small since these ratios are derived from transitions in the same 
spectral region.
While the continuous opacity may be affected more strongly by electron
scattering at the Pb~\textsc{i} transition in the ultraviolet, 
the abundances derived from this line generally agree with those
derived from the redder lines in individual stars in our sample.

We compare our derived Pb, Th, and $^{12}$C/$^{13}$C for four
stars with previous high-resolution analyses in Table~\ref{comparetab}.
Pb abundances have been derived previously for only two of the five stars 
in which we derived a Pb abundance, and in both cases we agree
within the uncertainties.
Our Th abundances in \bdcowan, \cssneden, \cshayek, and \hehayek\ 
are in good agreement with previous studies.
The previously reported Th abundances in \mbox{HD~6268}, \mbox{HD~115444}, 
and \mbox{HD~186478}
exhibit a great degree of scatter, but our abundances are consistent
with these measurements.

\begin{figure}
\epsscale{1.15}
\plotone{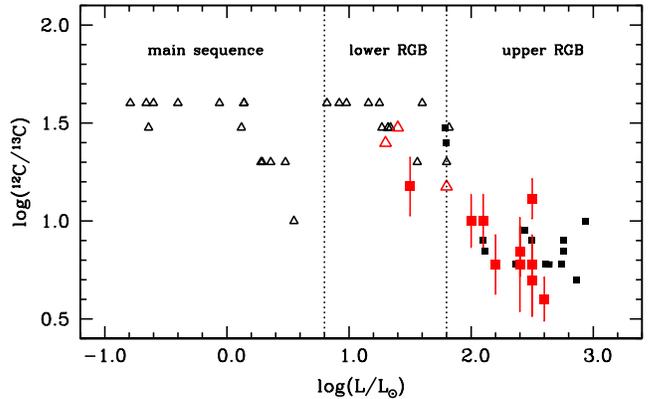} 
\caption{
\label{cratio}
$^{12}$C/$^{13}$C ratios as a function of luminosity.  
Filled black squares represent measurements from \citet{gratton00}
and open black triangles represent lower limits for stars with
[Fe/H]~$<-1.0$.
Filled red squares and open red triangles indicate our measurements
and lower limits, respectively.
Evolutionary classifications from \citet{gratton00} are indicated.
}
\end{figure}

Figure~\ref{cratio} compares our $^{12}$C/$^{13}$C ratios with 
those derived in \citet{gratton00} for metal-poor stars on the main sequence, 
lower red giant branch (RGB), and upper RGB.  
The individual luminosities for stars in our sample are also shown
in Table~\ref{cratiotab}.
The majority of stars in our sample lie on the upper RGB and all have
$^{12}$C/$^{13}$C in good agreement with the \citet{gratton00} stars
at similar luminosities.
The dredge-up processes that moderate the decreasing $^{12}$C/$^{13}$C 
ratio with increasing luminosity have no effect on the \ncap\ abundances
in these stars.

To explore the systematic uncertainties present when mixing 
abundances from different studies, we have compared the 
\eps{Eu} abundance and \eps{La/Eu} ratio derived in the present study to 
\citet{honda04} (5~stars in common), 
\citet{sneden09} (3~stars), 
\citet{aoki08b} (2~stars), and
\citet{hayek09} (2~stars).
\citet{sneden09} used the same spectra, model atmosphere grid and 
parameters, and 
analysis code to derive abundances as we have, differing only in the
list of lines and the ``human element''
present when different investigators make the same measurement.
\citet{hayek09} used the same spectra and model atmosphere parameters
as we have, differing in all other components.
\citet{honda04} and \citet{aoki08b} 
have used the same grid of model atmospheres as we have,
but we have no other components of our analysis in common.
All five of these studies employed spectral synthesis techniques to 
derive the abundances of La and Eu.

We find negligible offsets with respect to \citet{sneden09},
$\Delta$\eps{Eu}~$=-0.03\pm0.03$ and $\Delta$\eps{La/Eu}~$=-0.01\pm0.02$.
We find moderate and significant offsets with respect to 
\citet{honda04}, \citet{aoki08b}, and \citet{hayek09} in $\Delta$\eps{Eu},
$-0.09\pm0.07$, $+0.14\pm0.13$, and $+0.08\pm0.01$~dex, respectively.
This is not unexpected given all of the different components that 
enter into the derivation of an elemental abundance.
What is surprising, perhaps, is that significant differences are also 
found when comparing the \eps{La/Eu} ratio, which should be largely 
insensitive to differences in the model atmosphere grid and parameters.
The differences are largest with respect to \citet{honda04} and
\citet{aoki08b}, $-0.11\pm0.02$ and $-0.10\pm0.01$, respectively;
a smaller difference is found when comparing with \citet{hayek09}, 
$+0.04\pm0.01$.  
This suggests, perhaps, that a significant source of the difference
arises from the lines used and the algorithm for reduction and 
continuum normalization of the stellar
spectrum, although one should be somewhat cautious about over-interpreting
these differences with only two stars in common between our study
and each of \citet{aoki08b} and \citet{hayek09}.

\begin{figure*}
\includegraphics[angle=270,width=7.0in]{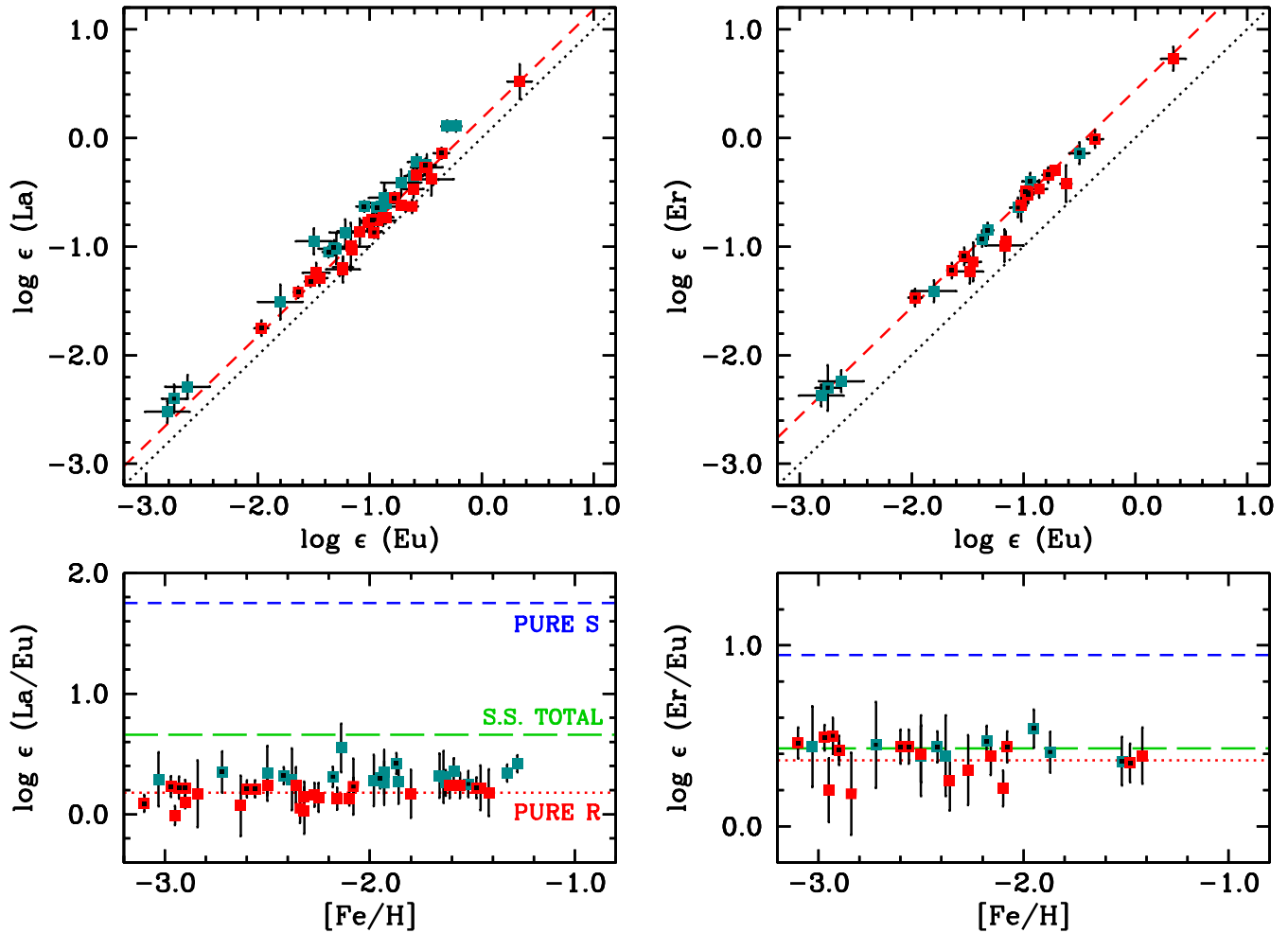} 
\caption{
\label{ratioplot1}
Comparison of the La/Eu and Er/Eu ratios in our sample.
Red squares represent measurements in stars with \eps{La/Eu}~$<+0.25$,
and blue squares represent measurements in stars with 
\eps{La/Eu}~$\geq+0.25$.
Black dots in the middle of these points signify measurements made
in the present study.
In the upper panels, the dotted black line represents a 1:1 ratio
of the elemental abundances and the red dashed line 
represents the mean ratio of the four ``standard'' $r$-only
stars in Table~\ref{means}.
In the lower panels, the red dotted line represents the 
pure-\rpro\ nucleosynthesis prediction 
(\citealt{sneden08}, with updates from Gallino),
the blue short dashed line represents the
pure-\spro\ nucleosynthesis prediction, and
the green long dashed line represents
the S.S.\ meteoritic ratio \citep{lodders03}.
}
\end{figure*}

\section{The $r$-Process Nature of Our Sample}
\label{rpro}

In order to correctly interpret the Th/Pb ratios
in these stars, it is important to demonstrate that the 
\rpro\ has been the only source of their \ncap\ material.
Even small contributions from the \spro\ will very easily 
bias the derived Pb abundances in \rpro\ enriched stars.
At low metallicity, an increase in the Pb abundance is
one of the earliest signatures of \spro\ nucleosynthesis.
Th can \textit{only} be produced in the \rpro\ and is
unaffected by \spro\ contributions.

The equivalence between the relative distributions of abundances 
for 56~$\leq Z \leq$~79 and the predicted 
\rpro\ contribution to these species' abundances in S.S.\ material,
seen in a growing number of metal-poor stars, 
is clear evidence that the \rpro\ has been the only significant 
source of \ncap\ material in these stars \citep{sneden08,sneden09}.
These stars include 
\bdcowan\ \citep{cowan02},
\cssneden\ \citep{sneden96,sneden03},
\cshill\ \citep{hill02},
\mbox{HD~115444} \citep{westin00},
\mbox{HD~221170} \citep{ivans06}, and
\hefrebel\ \citep{frebel07}.
\citet{sneden09} have recently remeasured and/or updated the 
rare earth (i.e., 57~$\leq Z \leq$~72) abundances in five of these stars.
Since the Th and U abundances of \cshill\ are enhanced relative to the
rare earths, for now we will exclude this star from the set of standards.
A complete chemical analysis of \hefrebel\ is underway (A.\ Frebel et al.,
in preparation).
We accept the remaining four stars
as the template for ``standard'' \rpro\ enrichment.

In the \rpro, the La/Eu ratio is $\approx$~1.5, whereas 
in the \spro\ the La/Eu ratio is $\approx$~56.  
In S.S.\ material, about 69\% of the La originated in the \spro,
whereas only about 5\% of the Eu originated in the \spro\
(\citealt{sneden08}, with updates from Gallino).
Furthermore, La and Eu are two elements with multiple 
absorption features in the spectra of metal-poor stars, and the $\log(gf)$ 
values for these transitions are well-known, so their abundances can be
derived with minimal line-to-line scatter.
The La/Eu ratio is an excellent discriminant of the relative
amounts of $s$- and \rpro\ material present in these stars.
In Figure~\ref{ratioplot1} we show the La/Eu ratios as a function of
[Fe/H] for our entire sample.
For comparison, the pure \spro\ and pure \rpro\ 
nucleosynthesis predictions for the La/Eu ratio 
and the S.S.\ ratio are also shown.

It is clear that the measured La/Eu ratios
in these stars lie close to the pure \rpro\ predictions---but how close?
The mean \eps{La/Eu} ratio for the four standard stars is 
$+0.18 \pm 0.03$ ($\sigma = 0.06$), ranging from 
$+$0.09 (\cssneden) to $+$0.23 (\bdcowan).  
We conservatively estimate that
any star with \eps{La/Eu}~$\geq +0.25$ has 
a non-negligible amount of \spro\ material present.
Assuming \eps{La/Eu}$_{{\rm pure-}r} = +0.18$ and
\eps{La/Eu}$_{{\rm pure-}s} = +1.75$, this limit identifies 
stars with no more than $\approx$~0.5--1.0\% of their \ncap\ material
originating in the \spro.
This explicitly assumes that the four \rpro\ standard stars 
contain no amount of \spro\ material.
The range and uncertainties of the La/Eu ratios in these stars 
set the limit of our ability to determine this percentage.
Even if the actual pure \rpro\ La/Eu ratio was 0.1~dex lower
than our mean---roughly equivalent to the La/Eu ratio in 
\cssneden, which has the lowest La/Eu ratio of any of our standard
stars---this limit would still represent only a 1.1\% contribution
from the \spro.
According to this definition, 27 stars in our sample 
have been enriched by only the \rpro, which we refer to as the
``\rpro-only'' sample for the remainder of this paper.

Elemental abundances for \ncap\ elements are usually sums over
multiple naturally-occurring isotopes of these species. 
\citet{sneden02}, \citet{aoki03}, \citet{lundqvist07}, and \citet{roederer08}
have shown that several of the stars in our \rpro-only sample 
have samarium  and Eu isotopic mixes 
consistent with \rpro\ nucleosynthesis.
This lends further credibility to our assertion that the \ncap\ material
in stars in our sample originated only in the \rpro.

\begin{figure}
\epsscale{1.15}
\plotone{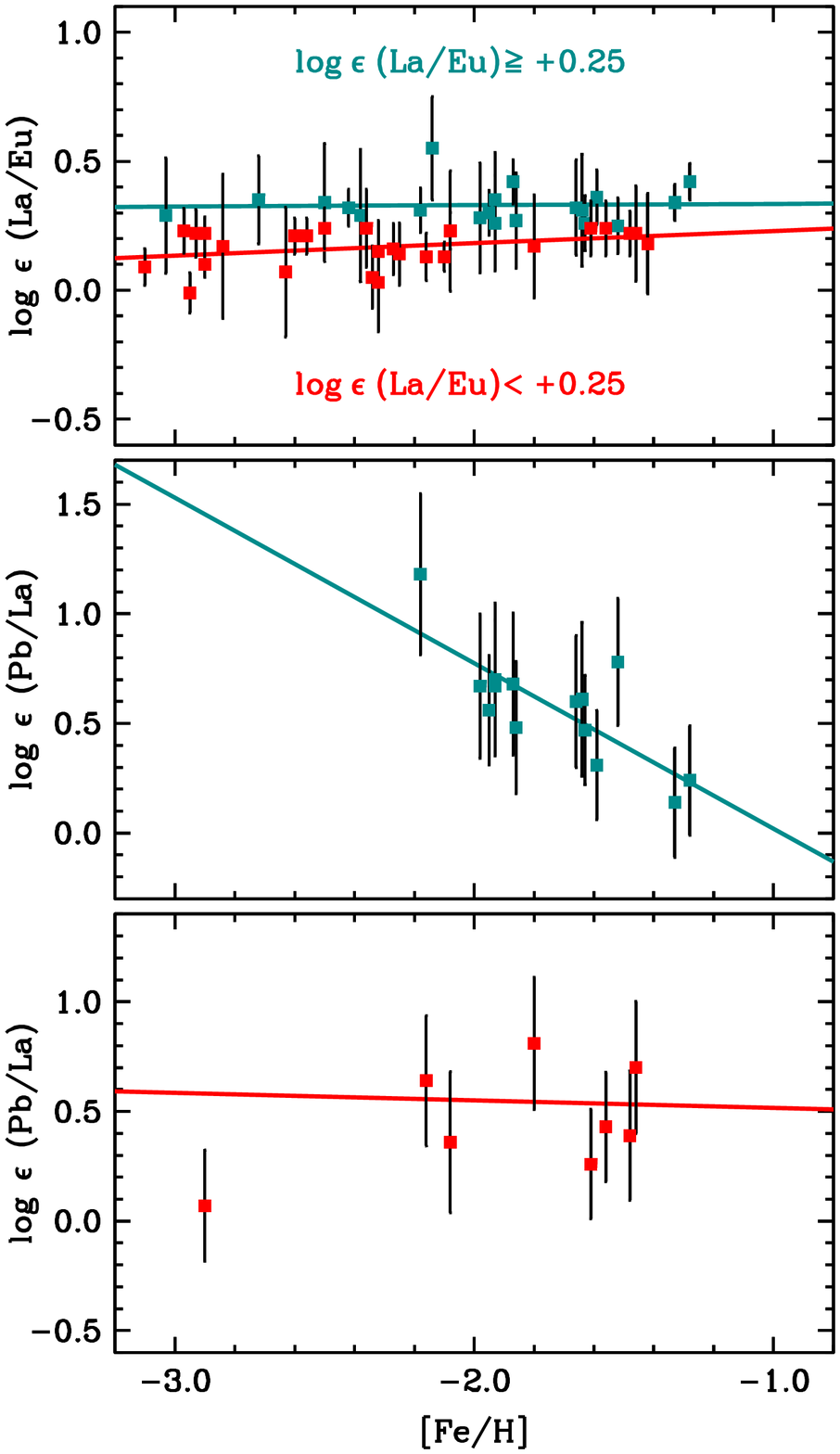}
\caption{
\label{spropb}
La/Eu and Pb/La ratios in stars with a hint of \spro\ enrichment
and stars with only \rpro\ enrichment.
Blue squares represent stars with \eps{La/Eu}~$\geq+0.25$, 
while red squares represent stars with \eps{La/Eu}~$<+0.25$.
The lines represent least-squares fits to the points.
The top panel demonstrates that neither of these two groups of stars 
exhibits any significant slope in La/Eu with [Fe/H].
It is clear that the Pb/La ratio increases with decreasing metallicity
in stars with a small \spro\ contribution (middle panel), 
while no such increase is
discernible in stars with \eps{La/Eu}~$<+0.25$ (bottom panel).
The star \cshill\ ([Fe/H]~$=-2.90$, \eps{Pb/La}~$=+0.07$) 
has been excluded from the fit in the bottom panel; see Section~\ref{nucleo}
for details.
}
\end{figure}

Now consider the stars in our sample with just a small 
amount of \spro\ material, those with \eps{La/Eu}~$\geq +0.25$.
The La/Eu, Pb/Eu, and Pb/La abundances for these stars are shown in
Figure~\ref{spropb}.  
While the La/Eu ratio shows no evolution with [Fe/H], 
the Pb/La ratio displays a marked increase with decreasing metallicity. 
This demonstrates that when the \spro\ operates at low
metallicity a relatively large amount of material accumulates at
the \third\ \spro\ peak due to the higher neutron-to-seed ratio.
Furthermore, this effect is noticeable in stars where only 
$\approx$~2.0\% of the \ncap\ material originated in the \spro\ 
(as determined from the La/Eu ratios of the stars with Pb/La ratios
displayed in the middle panel of Figure~\ref{spropb}).
In contrast, in stars with \eps{La/Eu}~$<+0.25$ the Pb/La ratio 
displays no trend with metallicity.
The combination of these two facts reinforces our assertion that
by choosing \eps{La/Eu}$_{r} < +0.25$ we have identified a sample of 
stars that are free of any detectable traces of \spro\ enrichment.

\begin{figure*}
\includegraphics[angle=270,width=7.0in]{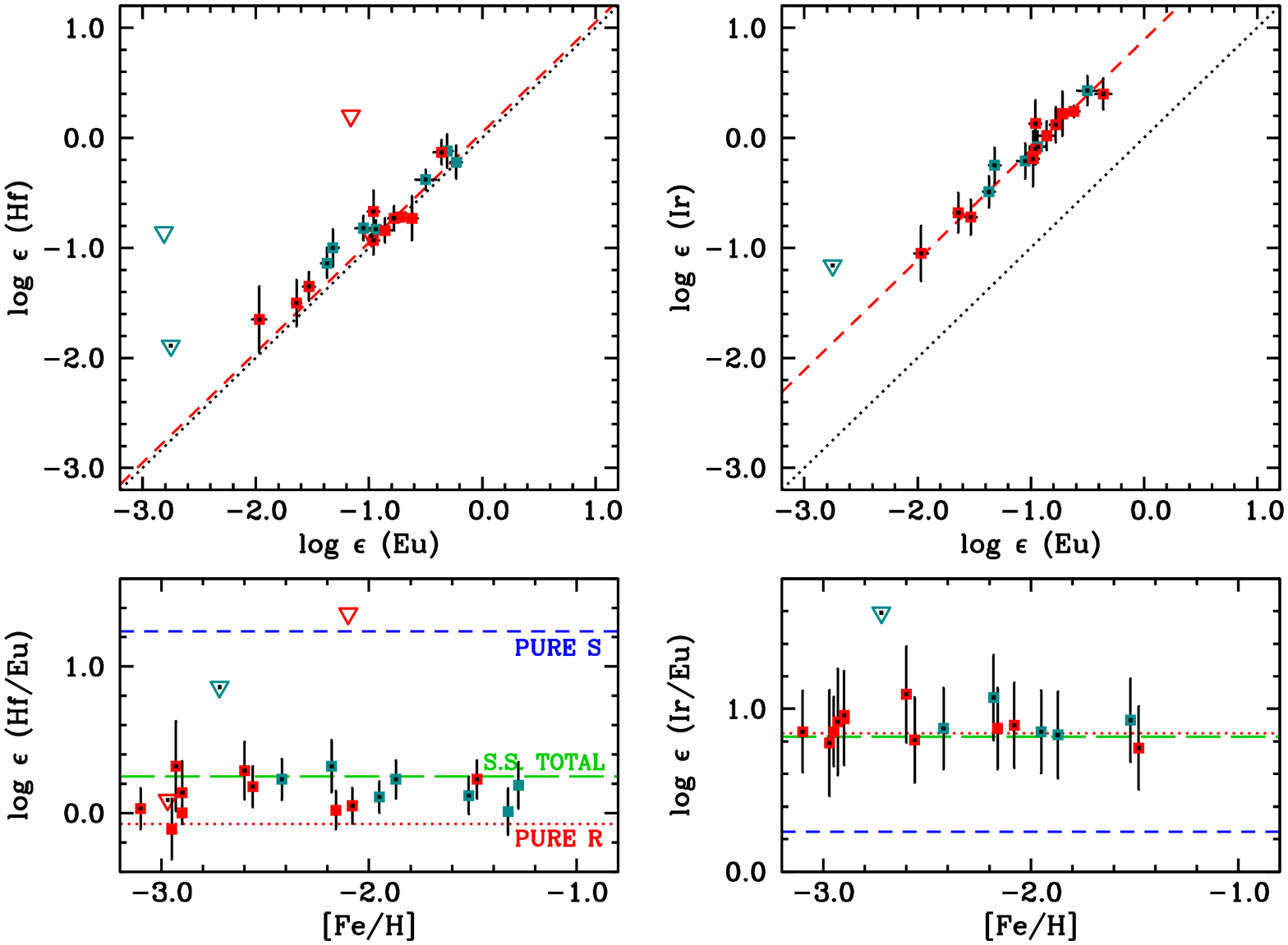} 
\caption{
\label{ratioplot2}
Comparison of the Hf/Eu and Ir/Eu ratios in our sample.
Symbols are the same as in Figure~\ref{ratioplot1}.
Downward facing triangles represent upper limits.
}
\end{figure*}

In Figures~\ref{ratioplot1} and \ref{ratioplot2}
we examine the Er/Eu, Hf/Eu, and Ir/Eu ratios.
By comparing these abundance ratios to those found in the four 
\rpro\ standard stars---whose compositions have been analyzed in 
excruciating detail for all of the \ncap\ species with 
accessible transitions---we can further characterize and establish
the \rpro-only nature of our sample. 
Stars with a hint of \spro\ material (identified by their La/Eu ratios)
are marked in blue, and stars with only \rpro\ material are marked in red.
The $r$/Eu ratios in stars with an \rpro-only signature are consistent 
with a single value.
The mean stellar Hf/Eu ratio differs somewhat
from the predicted \rpro-only ratio, suggesting
that the predicted S.S.\ breakdown for Hf may need minor revisions
(\citealt{lawler07}, \citealt{sneden09}; see further discussion
in \S~\ref{stellarages}).
The mean ratios for the \rpro-only stars are listed in Table~\ref{means}.
The $r$/Eu ratios are constant for La ($A =$~139),
Er ($A =$~162--170), Hf ($A =$~174--180), and Ir ($A =$~191--193),
extending through the entirety of the rare earths and to the
\third\ \rpro\ peak.

\begin{figure*}
\includegraphics[angle=270,width=7.0in]{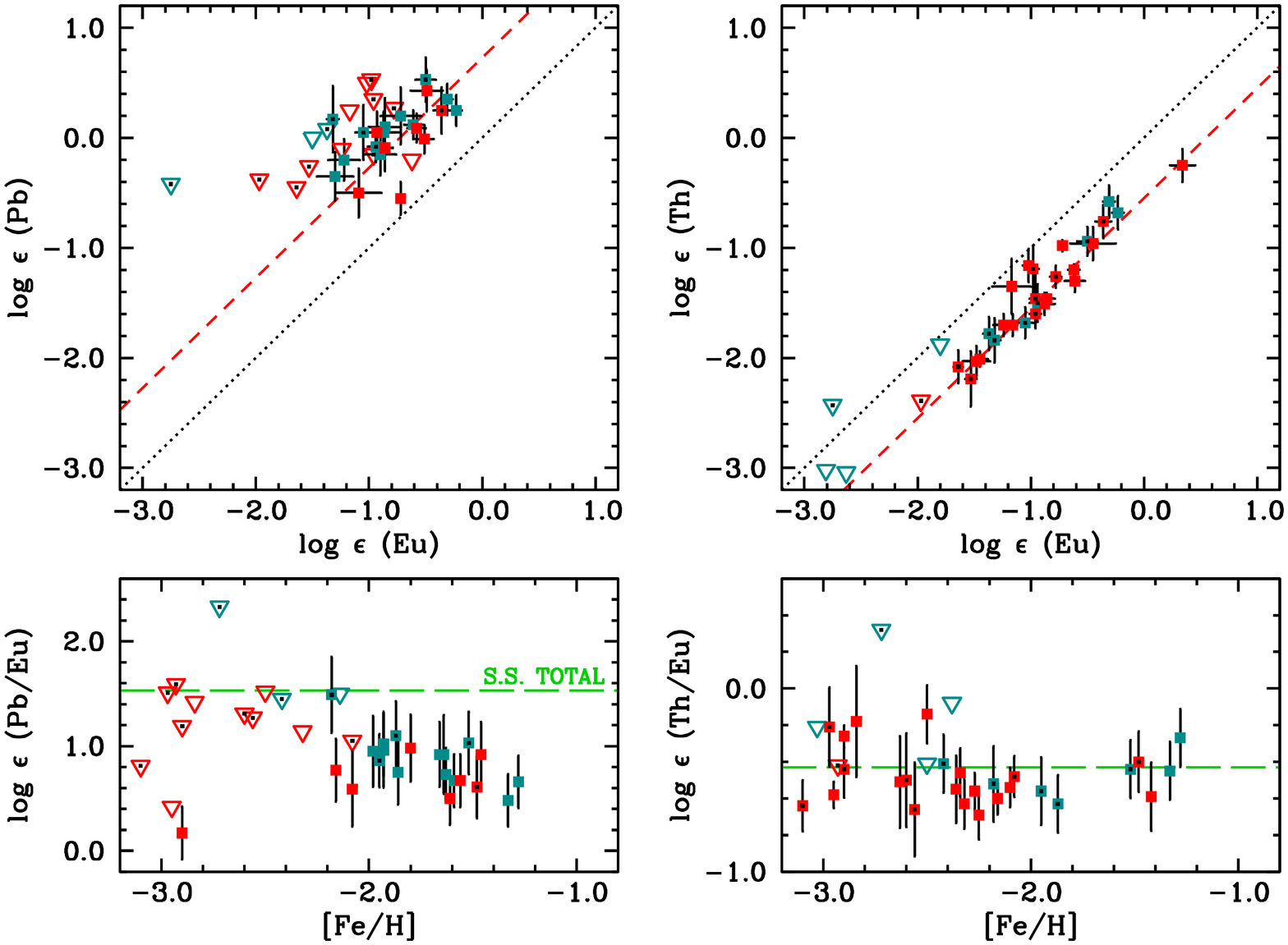} 
\caption{
\label{ratioplot3}
Comparison of the Pb/Eu and Th/Eu ratios in our sample.
Symbols are the same as in Figures~\ref{ratioplot1} and \ref{ratioplot2}.
}
\end{figure*}

Figure~\ref{ratioplot3} displays the Pb/Eu and Th/Eu ratios.
The Th/Eu ratio for nearly all
stars is consistent with a single value over the entire metallicity 
range of the measurements.  
This remarkable correlation is not just seen in the field stars, but 
also in globular clusters and one star in a dSph system.
Despite our efforts to detect Pb in low-metallicity stars enriched 
by the \rpro, only one convincing detection exists below [Fe/H]~$=-2.2$,
\cshill\ ([Fe/H]~$=-2.9$, \citealt{plez04}).
\hefrebel\ ([Fe/H]~$=-2.95$, \citealt{frebel07}) 
has a Th/Eu ratio consistent with the standard
\rpro-only stars, yet the Pb upper limit derived by \citet{frebel07}
indicates that the Pb in this star lies at least 0.3~dex below the \rpro\ value
seen in the \rpro-enriched stars with $-2.2 <$~[Fe/H]~$<-1.4$.
\cssneden\ ([Fe/H]~$=-3.1$, \citealt{sneden03})
has a Pb upper limit that is nearly identical to the 
Pb abundance found in the stars with $-2.2 <$~[Fe/H]~$-1.4$.
These three stars have approximately the same metallicity and 
high levels of \rpro\ enrichment
([Eu/Fe]~$= +1.6$, $+$1.8, and $+$1.6, respectively).
For the majority of metal-poor stars with known Pb and Th abundances, the 
\rpro\ pattern appears to continue to the actinides as well.
A few notable exceptions are discussed in Section~\ref{nucleo}.

\section{Nucleosynthesis of Pb and Th in the $r$-process}
\label{nucleo}

Four stars in the \rpro-only sample have Th/Eu ratios elevated by 
$\sim$~0.3--0.4~dex relative to the other stars in the \rpro-only sample:
\cshonda, \cslai, \cshill, and \hehayek.
This ``actinide boost'' \citep{schatz02} 
describes the enhanced Th (and, in \cshill, U)
abundance ratio(s) relative to the rare earth elements.
All four stars have metallicities in the range $-3.0 <$~[Fe/H]~$< -2.5$
and have high levels of \rpro\ enrichment
([Eu/Fe]~$\geq +0.85$, including three
with [Eu/Fe]~$>+1.2$), but several other stars with no actinide boost
also have similar levels of \rpro\ enrichment (e.g., \cssneden\ and
\hefrebel).
The distinction between these four stars and the remainder of the 
\rpro\ sample is rather clean, with no stars having 
$-0.40 <$~\eps{Th/Eu}~$< -0.26$.

\begin{figure}
\epsscale{1.15}
\plotone{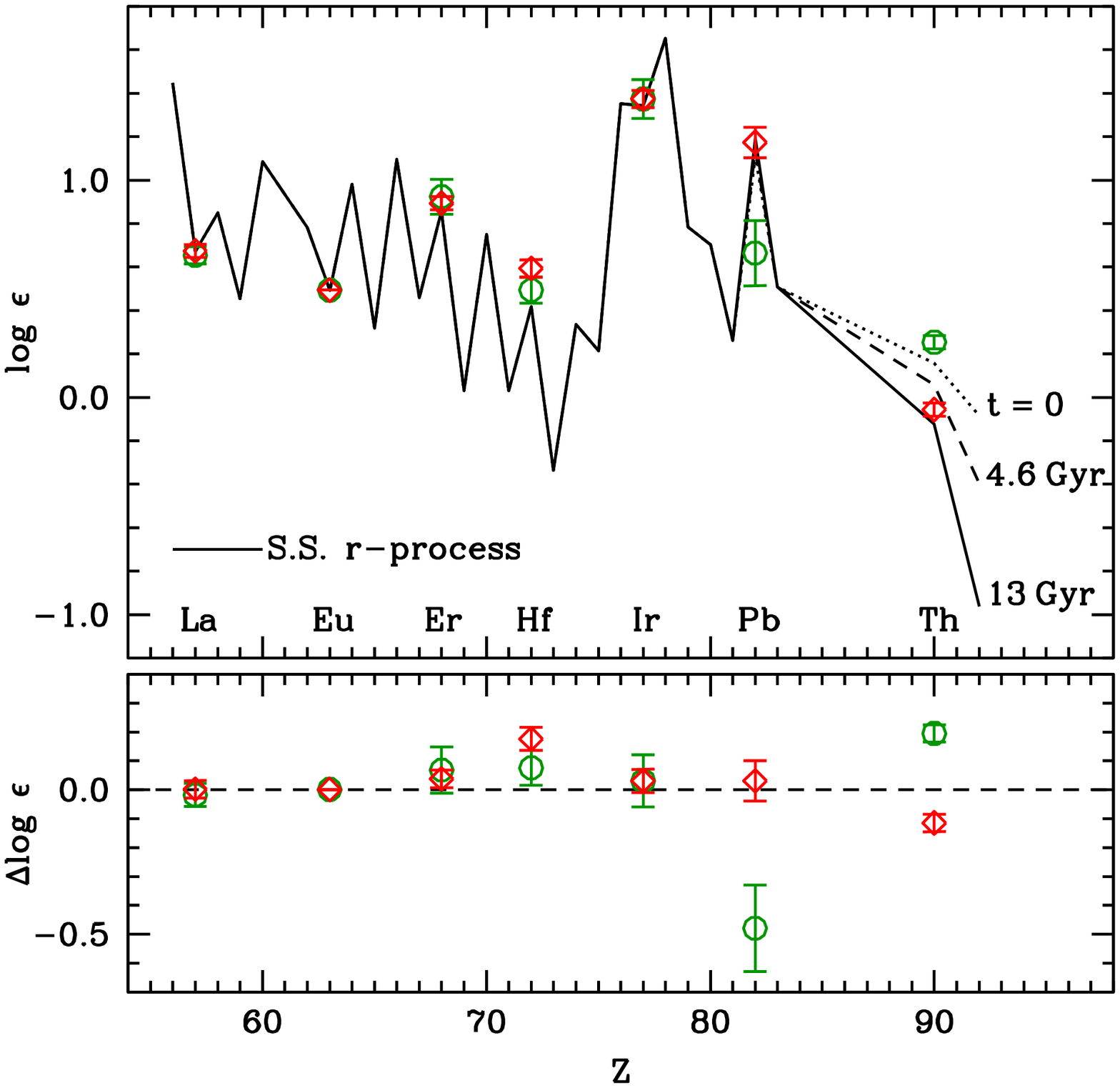} 
\caption{
\label{solarplot}
Comparison of the mean ratios for stars exhibiting a pure \rpro\
signature (\eps{La/Eu}~$<+0.25$; red diamonds)
with the four ``actinide boost'' stars (green circles).
The predicted S.S.\ \rpro ``residual'' abundance pattern is shown
for reference. 
The abundances are normalized at Eu.
Different decay ages are indicated by the dotted ($t = 0$~Gyr),
dashed ($t = 4.6$~Gyr), and solid ($t = 13.0$~Gyr) lines.
Any deviations in the abundances of the ``actinide boost'' stars 
from the ``standard'' \rpro\ stars 
clearly occur only after the \third\ \rpro\ peak.
}
\end{figure}

In Figure~\ref{solarplot} we compare the mean abundances of these
four stars to those of the other \rpro-only stars 
(also see Table~\ref{means}).
The S.S.\ \rpro\ pattern is shown for reference.
The abundances are normalized to the S.S.\ Eu abundance.
The mean La, Er, and Ir abundances between the two groups are identical;
the mean Hf abundance in stars with the actinide boost is 
determined from only one star, so we do not regard the 0.10~dex 
discrepancy as significant.  
The mean Pb abundance in stars with the actinide boost is also derived from
only one star, but it differs from the mean of the standard \rpro-only
stars by 0.51~dex.
The abundances of the stars with the actinide boost are identical to the 
abundances of the stars without an actinide boost 
from the rare earth domain to the \third\ \rpro\ peak; 
any differences in the nucleosynthetic process(es) that produced 
their \ncap\ material affect---at most---only the region beyond
the \third\ \rpro\ peak.

\begin{figure}
\epsscale{1.00}
\includegraphics[angle=270,width=3.4in]{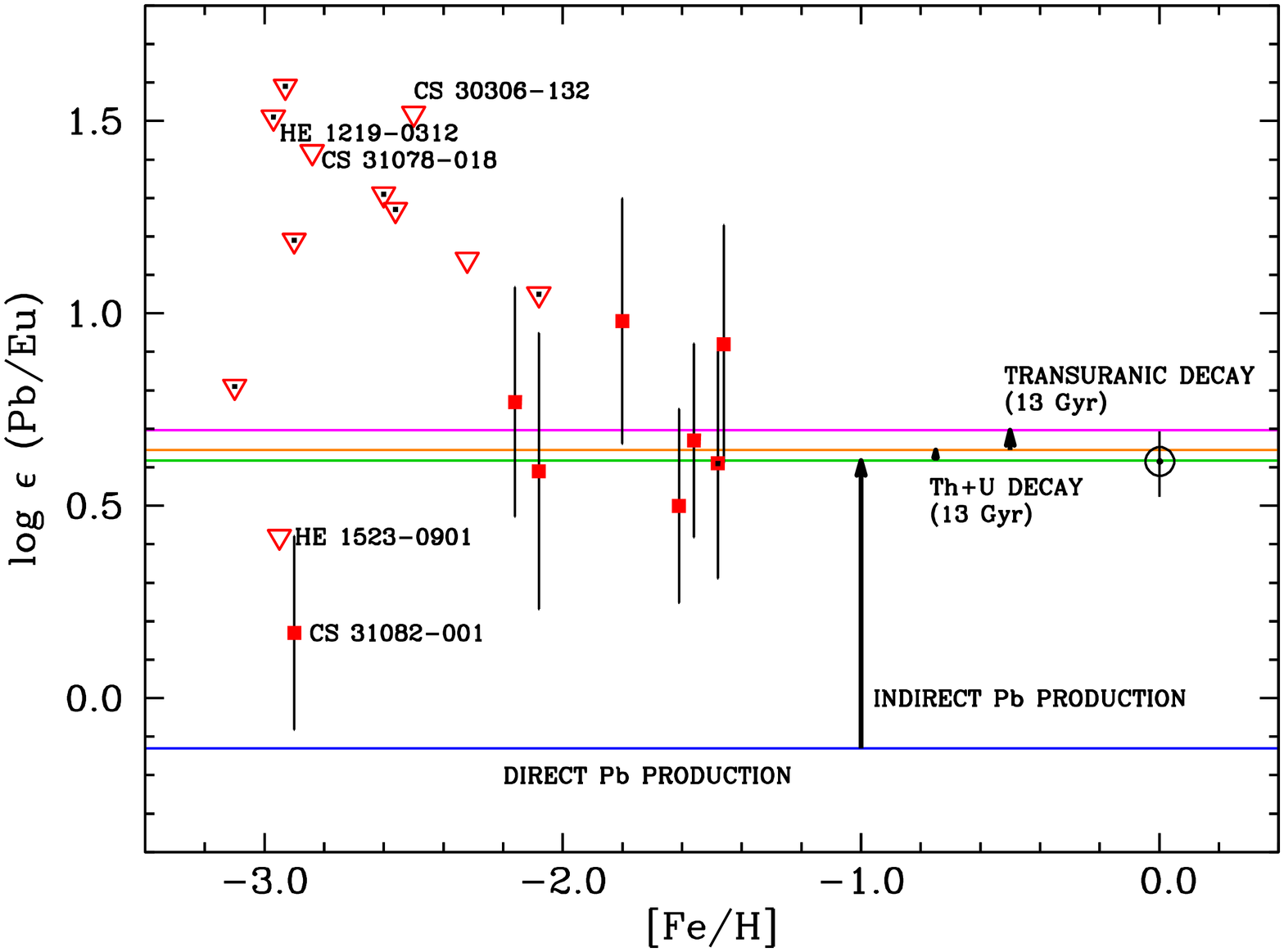} 
\caption{
\label{pbeuplot}
Comparison of \rpro-only stellar Pb/Eu ratios with Pb abundance predictions.
Symbols are the same as in Figure~\ref{ratioplot2}.
The Solar \rpro\ ratio is indicated by the $\odot$.
Predicted Pb/Eu ratios are indicated by horizontal lines, representing
contributions to the present-day Pb abundance via direct production
of the $^{206,207,208}$Pb isotopes; indirect Pb production via 
$\alpha$ and $\beta$ decays from nuclei with 210~$\leq A \leq$~231
and $A=$234
shortly after the termination of the \rpro\ event; 
13~Gyr of decay from nuclei that formed along the isobars of 
$^{232}$Th, $^{235}$U, and $^{238}$U; and 13~Gyr of decay
from the fraction of these three isotopes produced indirectly 
from the $\alpha$ and $\beta$ decay of transuranic nuclei shortly after
the \rpro\ event has shut off.
The excellent agreement of the predicted total Pb/Eu ratio with the 
stellar values (with the exceptions of \cshill\ and \hefrebel)
implies that fission losses from the region between 
Pb and Th are not significant.
}
\end{figure}

In Figure~\ref{pbeuplot} we show Pb/Eu ratios for the \rpro-only sample.
These stellar ratios are compared with our predictions, 
made using the classical waiting-point 
assumption---defined as an equilibrium condition between 
neutron captures and photodisintegrations---for the \rpro, 
employing the Extended Thomas-Fermi nuclear mass model
with quenched shell effects far from stability (i.e., ETFSI-Q, 
\citealt{pea96}).
Although this approach makes the simplifying 
assumptions of constant neutron number density and temperature 
as well as instantaneous nuclear freezout, the equilibrium model
calculations reproduce the S.S.\ abundances well;
see, e.g., \citet{kratz93}, \citet{cowan99}, \citet{freiburg99}, 
\citet{pfeiffer01}, and \citet{kratz07a}. 
These calculations are model-independent and look
only at the astrophysical conditions that lead to the \rpro\ and
reproduce the S.S.\ \rpro\ abundances. 
These waiting point calculations have been confirmed by more detailed  
dynamic (i.e., non-equilibrium) network calculations \citep{farouqi09}.
Our approach can be considered reliable 
{\it only} if we achieve a ``consistent''
picture---meaning that the abundances are solar---with 
logical astrophysical assumptions for the three heaviest 
\rpro\ ``observables;'' i.e., the
3rd peak, the Pb-Bi spike and the Th, U abundances. 

The specific 
calculations employed here assume a weighted range of neutron number densities
(from  10$^{23}$ to 10$^{30}$ cm$^{-3}$) and are designed to 
reproduce the total \rpro\ isotopic S.S.\ abundance pattern
(see \citealt{kratz07a} for more details). The best agreement with this
abundance pattern, as well as with the elemental abundances in 
metal-poor halo stars, is obtained by employing the nuclear mass 
predictions---many of the nuclei involved in this process are too 
short-lived to be experimentally measured at this time---of the 
ETFSI-Q nuclear mass model.
We employed 
the $\beta$-decay properties from quasi-particle random-phase
approximation calculations for the Gamow-Teller 
transitions (see \citealt{mol90}; \citealt{mol97}) and
the first-forbidden strength contributions from the Gross theory
of $\beta$-decay (\citealt{mol03}). 
We find that a small number of individual neutron
density components (in the current calculations, 15~components) 
is necessary to fit the predicted \rpro\ abundances to the S.S.\ values, 
and we also assume a varying \rpro\ path related
to contour lines of constant neutron separation energies in the 
range of 4--2~MeV.  (This path is determined by the variations of
the neutron number density and the temperature).

The nuclear data for these calculations have been improved by
incorporating recent experimental results and improved theoretical
predictions.  
Analyses of the differences between measured and predicted 
nuclear parameters (e.g., $\beta$-decay properties) 
indicate considerable improvements over earlier attempts. 
This gives us confidence in the reliability of our
nuclear physics input to the \rpro\ calculations of the heavy
element region between the rare earth elements, 
via the \third-peak elements (Os, Ir, Pt),
the Pb and Bi isotopes, and up to Th and U.
In addition the excellent agreement between
these calculations and the S.S.\ isotopic and elemental abundances
suggests that this approach can reproduce the 
astrophysical and nuclear conditions appropriate for the \rpro\, despite 
not knowing the astrophysical site for this process \citep{kratz07a}.
For this paper the theoretical 
predictions have been normalized to the \rpro\ component (95\%)
of the S.S.\ Eu abundance \citep{lodders03}.

The predicted Pb abundance shown in Figure~\ref{pbeuplot} 
is broken into four components corresponding to its origin within
the \rpro.
``Direct production'' refers to Pb that is produced in the \rpro\ as
nuclei with $A=$~206, 207, or 208, each of which will $\beta^{-}$
decay directly to one of the stable Pb isotopes.
``Indirect production'' refers to Pb that is produced from the 
$\alpha$ and $\beta$ decay of nuclei with $210 \leq A \leq 231$
and $A=$~234 shortly after the termination of the \rpro\ (i.e., within a 
few~$\times10^{6}$ years).
``Th and U decay'' refers to Pb that originates from the 
decay of nuclei that were produced in the 
\rpro\ with $A=$~232, 235, or 238, which quickly $\beta^{-}$ decayed
to $^{232}$Th, $^{235}$U, and $^{238}$U, which have since decayed
to the stable Pb isotopes.
In Figure~\ref{pbeuplot} we show the Pb abundance after 13~Gyr.
``Transuranic decay'' refers to Pb that is produced from the decay of
$^{232}$Th, $^{235}$U, and $^{238}$U, but now considering the
fractions of the abundances of these three isotopes that 
were formed in the \rpro\ as nuclei with $A=$~236 and $A \geq 239$, which
followed $\alpha$ and $\beta$ decay chains 
to the long-lived Th and U isotopes. 

The abundance predictions for Pb, Th, and U, computed at time ``zero''
after all of the $\alpha$ and $\beta$ decays are complete 
($\sim 10^{7}$~years after the \rpro\ event) 
are listed in Table~\ref{pbtab}.
According to these predictions, the majority of the \rpro\ Pb abundance
derives from indirect production of short-lived nuclei between
Pb and Th (82\% at time ``zero'').
Only a small amount of the present-day Pb originated from the 
decay of the long-lived Th and U isotopes.
The predicted total Pb/Eu ratio, based upon our waiting-point \rpro\
calculations as described above, is in agreement with the derived
Pb/Eu ratios in all of the stars with $-2.2 <$~[Fe/H]~$<-1.4$,
independent of the amount of time since the \rpro\ nucleosynthesis event.
This implies that fission losses from nuclei with $210 \leq A \leq 231$
and 234 are not significant, otherwise the predicted total Pb abundance 
would be noticeably lower.
These results, based upon detailed fits that reproduce the total S.S.\ 
isotopic abundance distribution, 
are supported by recent dynamic calculations (see, e.g.,
\citealt{farouqi09}), which also indicate that the total amount of fission 
material in this mass range will be insignificant and not 
affect the Pb to Th and U region. 
An additional argument regarding the lack of any appreciable 
effects of fission comes from
recent fission barrier height calculations of M\"oller (2009, private
communication), who finds that any  
fissioning isotopes in the mass region 
206~$<A<$~232--238---if they exist at all---lie well beyond the
\rpro\ path in our model predictions (i.e., in isotopes more neutron-rich
than the \rpro\ path and closer to the neutron drip line).
Any fissioning isotopes would not normally
contribute to the abundances in the Pb region (even if they did not fission). 
Our results also indicate that any fissioning nuclei
in the mass region 232--238~$<A<$~250--256 make no significant contributions
to the stable heavy isotopes. 

We emphasize here that our approach has been to fit
the observed S.S.\ isotopic (and elemental) abundance distribution with 
the superposition of the neutron number density exposures that 
occur in the \rpro.
These calculations are further strengthened by 
utilizing the most comprehensive set of nuclear
data currently available---including experimentally measured and theoretically 
determined (published and unpublished ETFSI-Q) masses, 
$\beta$-decay half-lives, and \ncap\ rates.
Our normal procedure has been to globally fit the S.S.\ abundances,
particularly normalizing the fit to the \second\ and \third\ process 
peaks. In addition to employing these global fits 
we have also fine-tuned our theoretical calculations
to reproduce the complete S.S.\ \third\ abundance peak in the
mass range of Os ($A=$~186) to Hg ($A=$~204). 
It has long been argued that the production of the radioactive
elements extending upwards from Pb to Th and U and the heaviest 
stable elements in the \third\ \rpro\ peak are correlated 
\citep{cowan91,thielemann93,cowan99,kratz04,kratz07a}.
Only the correct reproduction
of the $A=$~195 abundance peak with its $N=$~126 (nuclear) 
bottle-neck behavior will guarantee that the extrapolated abundances
into the experimentally (completely) unknown \rpro\ trans-Pb region,
including exotic isotopes about
30--40 units away from the known ``valley of the isobaric mass parabolae,'' 
will be reliable.

The excellent agreement between the \rpro\ Pb abundances in stars 
with $-2.2 <$~[Fe/H]~$<-1.4$ and our predictions is encouraging.
Assuming the stellar Pb abundances are not seriously in error, we
currently lack a complete, self-consistent understanding of \rpro\ 
nucleosynthesis and enrichment for all low metallicity stars.
Both \cshill\ and \hefrebel\ have similar metallicities and levels of
\rpro\ enrichment.
Their U/Th ratios are similar, and
their Pb abundances may be similar to one another 
(yet different from more metal-rich \rpro-enriched stars).
Their Th/Eu ratios are different from one another,
yet the Th/Eu ratio in \hefrebel\ agrees with most other 
metal-poor stars, while \cshill\ does not.
(See further discussion in \citealt{frebel09}.)
A still larger set of \rpro-enriched stars with [Fe/H]~$< -2.2$ will be
necessary to characterize the abundance patterns of Pb, Th, and U
at low metallicity, and until that time (at least) our understanding
of the nucleosynthesis of the heaviest products of the \rpro\ 
will remain incomplete.

\section{Stellar Ages}
\label{stellarages}

When deriving stellar ages through nuclear chronometry, 
the measured ages reflect the ages of the 
actinides in these stars but not necessarily the ages of the stars themselves.
In addition, a few \rpro\ events may have
seeded the ISM from which these stars formed.
Presumably the stars formed shortly after the \rpro\ material
was created and so the derived age 
represents an upper limit to---but also a realistic estimate of---the 
stellar age.

U/$r$ ratios (where $r$ denotes a stable element 
produced in the same \rpro\ event as the U) have the 
strongest predictive power (0.068~dex per Gyr),
followed by U/Th (0.046~dex per Gyr), and
Th/$r$ (0.021~dex per Gyr).
These rates are only governed by the nuclear half-lives of the
radioactive isotopes, which have been measured to exquisite precision
as far as stellar nuclear chronometry is concerned.
Strictly speaking, the Pb/$r$ (or U/Pb, or Th/Pb) ratio is also a 
chronometer, since the Pb abundance increases with time as
the Th and U decay, but it loses sensitivity as time passes and 
most of the Th and U nuclei decay.
According to our model, 90\% of the total
increase in the Pb abundance has already occurred
in the first 5~Gyr after the \rpro\ event, and 
the total (i.e., $t = \infty$) Pb abundance is only 0.10~dex higher
than its time ``zero'' abundance.

Stellar abundance ratios are rarely known to better than
0.05--0.10~dex, limiting relative age determinations from one star
to another using a single Th/$r$ pair to a precision of $\sim$~3--5~Gyr.
To some degree this can be mitigated by employing multiple 
Th/$r$ pairs, but uncertainties in the production ratios and 
uncertainties arising from systematic effects (e.g., determination
of effective temperature) will limit the measurement of an
absolute age to no better than $\sim$~2--3~Gyr 
(see, e.g., \citealt{sneden00} or \citealt{frebel07}).

\begin{figure*}
\includegraphics[angle=270,width=7.0in]{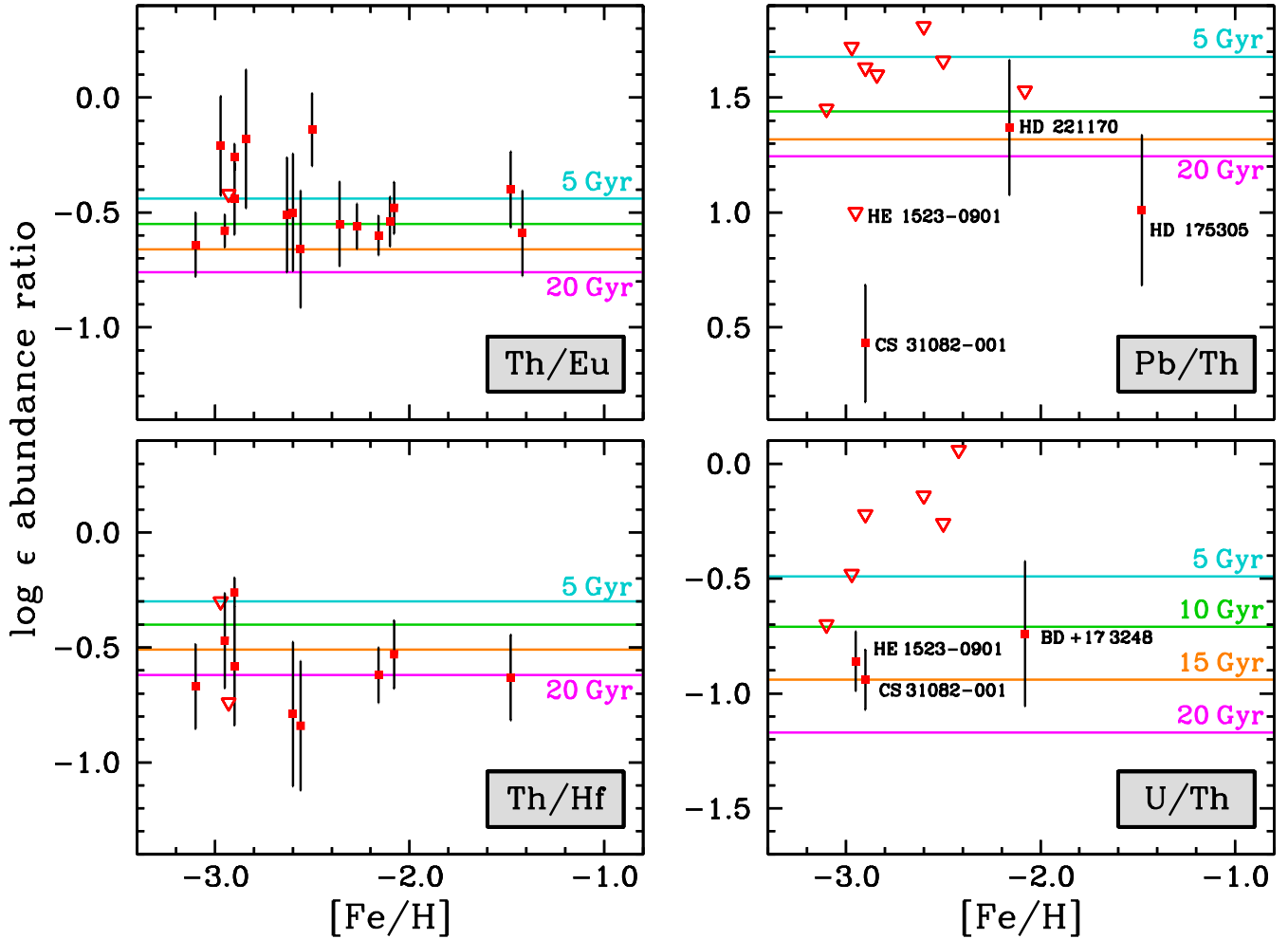} 
\caption{
\label{agesplot}
Four nuclear chronometer pairs.
Only stars with a pure \rpro\ signature are shown.
Symbols are the same as in Figure~\ref{ratioplot2}.
The horizontal lines indicate the ratios expected if a sample
of material had a given age, assuming the nucleosynthesis predictions
of \citet{kratz07a}.
The vertical scales cover the same number of decades on each panel,
though the ranges differ, to illustrate the relative measurement 
precision.
The shortcomings of the \rpro\ Hf predictions by our model are
evident in the poor match between the Th/Hf predictions and 
observations shown in the lower left panel.}
\end{figure*}

Figure~\ref{agesplot} illustrates these points by showing 
the ages derived for our \rpro-only sample from
Th/Eu, Th/Hf, Pb/Th, and U/Th chronometers.\footnote{
The element Hf has recently been suggested as
a promising, stable \rpro\ reference element. 
\citet{lundqvist06} and \citet{lawler07} have made 
new laboratory evaluations of Hf~\textsc{ii} transition 
probabilities, and there are indications that Hf may be formed at 
similar neutron densities to the \third\ \rpro\ peak elements
Os, Ir, and Pt \citep{kratz07a}.
These elements are difficult to reliably measure in stellar spectra 
obtained from ground-based facilities (except for Ir)
and can only be observed in their neutral states.
In other words, Hf may be the heaviest stable singly-ionized element 
whose production is closely linked to that of the actinides
and may be reliably and easily measured in metal-poor stars.
Our model predicts an \rpro\ Hf abundance of 0.0436 
(26\% of the S.S.\ value of 0.1699 for 
$A =$~174--180, \citealt{lodders03}).
The isotopic and elemental Hf \rpro\ abundance predictions by our model
are consistently smaller than those inferred from the S.S.\
\rpro\ residual method.
We caution that knowledge of the nuclear structure of 
atoms is incomplete for isotopes near $A \sim$~180, in the transition from 
the deformed rare earth isotopes to the spherical $N =$~126 magic shell.
The offset between the scaled S.S.\ and stellar \rpro\ Hf abundances
noticed by \citet{lawler07} and \citet{sneden09} suggests
that a larger fraction of the S.S.\ Hf abundance might be 
attributed to the \rpro.
That study adopted the S.S.\ Hf \rpro\ fraction (44\%) from 
\citet{arlandini99}, implying that the S.S.\ \rpro\ fraction is
even higher than 44\%.
If we reverse the problem and assume that the stellar Hf/Eu \rpro\ ratio
from \citet{sneden09} should match the S.S.\ Hf/Eu \rpro\ ratio, 
we estimate the \rpro\ fraction of Hf in the S.S.\ is $\approx$~70$\pm$10\%.}
Our results are consistent with the assertion that \rpro\ enrichment
began extremely early, within the first few Gyr after the big bang.
The horizontal lines indicating ages are determined based on the
production ratios given in \citet{kratz07a}.
Using a different set of production ratios (e.g., \citealt{cowan99},
\citealt{schatz02}, or those predicted by our stellar sample)
would only change the absolute scale of these age determinations 
by small amounts ($\sim$~2--4~Gyr; see \citealt{frebel07}).

This is the largest sample of Th/Eu ratios yet compiled for metal-poor 
halo stars.
The chronometer pairs in an individual star have limited ability
to predict its age, but the combined measurements from an 
ensemble of stars hold greater promise.
The majority of our sample suggests an old population
(the exception being the four stars with an actinide boost), 
and no trends with metallicity are apparent.
This is in agreement with model predictions 
for the age-metallicity relationship in metal-poor halo stars, 
where a rapid increase is found in the
mean metallicity to [Fe/H]~$\sim -$2.0 within the first 2~Gyr
(e.g., \citealt{argast00}).
If we divide the sample into two groups of stars---those 
with an actinide boost and those without---and 
assume a single age for each group,
we can derive reasonable estimates for the age of the \rpro-only standard
stars, as shown in Table~\ref{agestab}.
Assuming that the observed stellar ratios are independent
(which they clearly are not since all rely on Th),\footnote{
A change in the Th abundance by $\pm$0.05~dex will uniformly change the
derived age by $\mp$2.3~Gyr.
The uncertainty on the age derived from each chronometer pair
is computed by combining in quadrature the error of the mean
stellar ratio, an assumed 20\% uncertainty in the production ratio,
and the uncertainty in the abundances arising from uncertainties in the
stellar parameters.
For this final source of error, we consider the uncertainties 
derived by \citet{frebel07} for \hefrebel\ to be representative
of the uncertainties for an individual star in our sample, and we
reduce each of their uncertainties by the amount expected when
increasing the sample size from one star to several. 
}
We derive an age for the 
ensemble of standard \rpro-only stars of 15.2$\pm$2.1 ($\sigma$=4.6) Gyr.
Two ratios predict ages significantly greater than the 
age of the Universe, Th/La (20.4$\pm$4.2~Gyr) and 
Th/Hf (19.7$\pm$4.7~Gyr).
If we adopt the higher S.S.\ \rpro\ ``residual'' abundances for La
and Hf (e.g., \citealt{sneden08}, with updates from Gallino), 
the Th/La and Th/Hf ages decrease by 13 and 3~Gyr, respectively.

The four stars with an actinide boost reflect a very different age,
1.3$\pm$2.1 ($\sigma$=4.6) Gyr.
In \cshill, where abundances of both Th and U have been derived
\citep{cayrel01,hill02}, the U/Th ratio implies a reasonable age of 
14.2$\pm$2.4 ($\pm$sys) Gyr \citep{hill02}.
If we use the $^{238}$U/$^{232}$Th production ratio computed from the data
in Table~\ref{pbtab} and assume a 20\% uncertainty in this value, 
we derive an age of 15.1$\pm$4.3~Gyr in \cshill.
The low Pb abundance and high Th/$r$ ratios in this star 
cannot be accounted for simply by assuming a very young age; 
increasing the current Pb abundance in this star by the maximum expected
from the complete decay of the actinides ($\approx$0.10~dex) would still
leave it about 0.4~dex lower than the mean of the more metal-rich
\rpro\ enriched stars.
U has also been detected in two \rpro\ enhanced stars that do not
have an actinide boost, \bdcowan\ \citep{cowan02} and 
\hefrebel\ \citep{frebel07}, and the U/Th ages in these two stars
are consistent with the U/Th age in \cshill\ (see Figure~\ref{agesplot}).

\subsection{Comparison to Globular Cluster and dSph
Age Estimates from Other Methods}
\label{globulars}

Several stars in our sample are located in globular clusters or 
dSph systems, whose ages and star formation histories can be 
estimated based on other methods.
M15 is an old (13.2$\pm$1.5~Gyr, \citealt{mcnamara04}), metal-poor
([Fe/H]~$\sim-2.2$; \citealt{gratton04}) globular cluster.
\citet{sneden00} derived abundances of several \ncap\ elements
in three giants in this cluster.
The mean Th/Eu ratio for these three stars implies an age of
12.0$\pm$3.7~Gyr using our production ratio.
M92 is also an old (14.8$\pm$3~Gyr, \citealt{carretta00};
14.2$\pm$1.2~Gyr, \citealt{paust07}), metal-poor 
([Fe/H]~$\sim-2.3$; \citealt{gratton04}) globular cluster.
Using the Th/Eu ratio derived by \citet{johnson01} for one star
in this cluster, we find and age of 10.6$\pm$4.7~Gyr.
\citet{aoki07} derived the Th/Eu ratio in one star in 
UMi, which implies an age of 12.0$\pm$6.5~Gyr.
This result is consistent with 
the earlier finding that this system experienced only one significant 
episode of star formation at early times ($\gtrsim$~11~Gyr ago;
\citealt{dolphin02}).
In all three cases, the ages for these systems derived from 
nuclear chronometry and other independent methods agree
within the uncertainties.

\section{Implications for Nucleosynthesis in the Early Galaxy}
\label{earlygalaxy}

One of the more remarkable aspects of the abundance ratios for \rpro-only
stars in Figures~\ref{ratioplot1}, \ref{ratioplot2}, and \ref{ratioplot3}
is the wide range of Eu and Fe abundances covered by these correlations.
The $r$/Eu ratios are constant 
over a very wide range of absolute \rpro\ enrichment, 
roughly 2.4~dex or a factor of 250.  
All of the [$r$/Eu] abundance ratios in these figures
(with the noted exception of the [Pb/Eu] and [Th/Eu] ratios in the
stars with an actinide boost)
are unchanged over the metallicity range 
$-3.1 \leq$~[Fe/H]~$\leq -1.4$.
For these stars, the [Eu/Fe] ratios are always supersolar, but they
vary widely, from $+$0.31 (\mbox{M92~VII-18}) to 
$+$1.82 (\hefrebel).
This wide dispersion in \ncap\ abundances at low metallicities
has been previously noted by many investigators, including 
\citet{gilroy88} and \citet{burris00}.
Several stars with $+0.3 <$~[Eu/Fe]~$<+0.5$ exhibit an \rpro-only 
signature, which reveals---as might be expected---that a pure
\rpro\ pattern can be found even in small amounts of \rpro\ enrichment.
Furthermore, this same pattern is observed in stars in the Galactic
halo, several globular clusters, and one dSph system.
Taken together, these facts are strong evidence for the universal nature
of the main \rpro\ for species with $Z \geq 56$ (as constrained by 
observations), since
stars with \eps{Eu}~$\sim +0.4$~dex certainly are comprised of the 
remnants of many more supernovae than stars with 
\eps{Eu}~$\sim -2.0$.

From an analysis of the La/Eu ratio in a sample of 159 stars
with $-3.0 <$~[Fe/H]~$<+0.3$, \citet{simmerer04} found stars with
[Fe/H]~$= -2.6$ exhibiting signatures of the \spro, while other stars
as metal-rich as [Fe/H]~$=-1.0$ showed little evidence of any \spro\ material.
By the definition we adopt in Section~\ref{rpro}, \eps{La/Eu}$_{r} < +0.25$,
processes other than the \rpro\ (e.g., the so-called weak \rpro) must be 
responsible for some \ncap\ 
material at even lower metallicity. 
\citet{simmerer04} also found 
incomplete mixing of both \rpro\ and \spro\ material
in stars with [Fe/H]~$>-1.0$.
We find that the gas from which these stars formed 
was inhomogeneous at metallicities as high
as [Fe/H]~$=-1.4$, the metal-rich limit of our sample;
here, several stars show no evidence of any contributions from the \spro.
Furthermore, \citet{roederer09} found no preferred kinematic signature
for stars with pure \rpro\ or \spro\ enrichment patterns,
with these patterns extending over a wide metallicity range 
of $-3.0 <$~[Fe/H]~$<-0.4$.
This reinforces the notion that \ncap\ enrichment at low metallicity
is likely a very localized phenomenon that results in a large 
distribution of \ncap\ abundances.

The range of absolute \rpro\ enrichment noted above includes 18 stars in M15. 
(Three stars from \citealt{sneden00} are shown in Figure~\ref{ratioplot1};
additional stars from \citealt{sneden97} and \citealt{otsuki06}, 
who did not report Th abundances, are not shown.)
These 18 stars in a single globular cluster show no change in 
their Ba/Eu or La/Eu ratios despite 
a change in the absolute Eu enrichment level by 0.9~dex,
a factor of $\approx$~8; their Fe abundances differ by less than 0.2~dex,
a factor of $\approx$~1.5.
\citet{sneden97} found no correlation between these \ncap\ enrichment
patterns in M15 and the signatures of deep mixing commonly observed
in globular cluster stars, indicating a primordial origin.
This enrichment pattern resembles that of the Galactic halo, 
but it is not obvious why other clusters enriched by \rpro\ material
(with or without significant contributions from the \spro)
fail to show similar star-to-star dispersion.

\section{Conclusions}
\label{conclusions}

We have identified a sample of 27 stars with 
$-3.1 <$~[Fe/H]~$< -1.4$ that have been enriched by the
\rpro\ and show no evidence of \spro\ enrichment.
We confront \rpro\ nucleosynthesis predictions for Pb and Th
with measurements (or upper limits) in our stellar sample.
We use these very heavy isotopes located near the 
termination points of $s$- and \rpro\ nucleosynthesis 
to better understand the physical nature of the \rpro\ 
and the onset of nucleosynthesis in the early Galaxy.
Our major results can be summarized as follows.

Stars with \eps{La/Eu}~$<+0.25$ (our ``\rpro-only'' sample,
where more than $\approx$~99\% of the \ncap\ material originated in the \rpro) 
show no evolution in their Pb/La ratio over the metallicity range 
$-2.2 <$~[Fe/H]~$< -1.4$.
In contrast, stars with \eps{La/Eu}~$\geq +0.25$
(those with just a dusting of \spro\ material on top of
\rpro\ enrichment)
show a significant increase in Pb/La with decreasing metallicity.
This emphasizes the effect of the higher neutron-to-seed ratio 
that occurs in low metallicity \spro\ environments and 
overproduces nuclei at the termination of the \spro\ chain
relative to lighter \spro\ nuclei.
This effect is noticeable in stars where only 
$\approx$~2.0\% of the \ncap\ material originated in the \spro.

All stars in our \rpro-only sample have constant abundance ratios
among elements surveyed in the rare earth domain and the 
\third\ \rpro\ peak (La, Eu, Er, Hf, and Ir), and these
abundance ratios are equivalent to the scaled S.S.\ \rpro\ distribution.
These ratios are identical in stars with
a so-called actinide boost and stars without.
For stars with an actinide boost, our observations demonstrate that 
any nucleosynthetic deviations from the main \rpro\ 
affect---at most---only the elements beyond the \third\ $r$-process peak 
(Pb, Th, and U).

We find very good agreement between the Pb abundances in our 
\rpro-only stars and the Pb abundances predicted by our 
classical ``waiting-point'' 
\rpro\ model.
In these computations a superposition
of 15 weighted neutron-density components in the range 
23~$\le$~log~n$_n$~$\le$~30 
is used to successfully reproduce both the S.S.\ isotopic
distribution and the heavy element abundance pattern
between Ba and U
in the low-metallicity halo stars.
Our calculations indicate that fission losses are 
negligible for nuclei along the \rpro\ path between Pb and the
radioactive isotopes of Th and U.
In light of this agreement, we currently have
no viable theoretical explanation for the low Pb abundance in \cshill.

With the exception of the Pb and Th in stars with an actinide boost,
the $r$/Eu ratios in our \rpro-only sample 
are constant over a wide range of metallicity
($-3.1 <$~[Fe/H]~$< -1.4$) and \rpro\ enrichment 
($-2.0 <$~\eps{Eu}~$< +0.4$ or $+0.3 <$~[Eu/Fe]~$< +1.8$).
This pattern is observed in field stars, several globular clusters, and
one dSph system.
As multiple supernovae will have contributed to the material in
stars at the highest metallicities and/or \rpro\ enrichments surveyed,
we regard this as strong evidence for the universal nature of the \rpro.

We have derived an age of 15.2$\pm$2.1~Gyr ($\sigma = 4.6$~Gyr) 
from several Th/$r$ chronometer pairs for an ensemble of 16 stars.
This is the largest set of Th/$r$ ratios yet compiled for metal-poor
halo stars.
Excluding the four stars with an actinide boost,
there is no relationship between age and metallicity
over the range $-3.1 <$~[Fe/H]~$< -1.4$.
While each stellar chronometer pair ratio argues for a
single common age for the ensemble of stars, 
the observations call for further refinement of some
production ratios, and the outlook for improving the precision
of the age measurement for a single star in the field is not hopeful.
For two globular clusters and one dSph system, 
we do find good agreement (within uncertainties $\sim$~3.5--5~Gyr)
between ages derived from the Th/Eu nuclear chronometer and ages 
derived from other methods.

\acknowledgments

We thank Satoshi Honda and David Lai for sending unpublished 
upper limit measurements and David Yong for providing 
additional abundance results.
We appreciate the helpful suggestions of the referee.
I.U.R.\ also thanks Jim Truran for an invigorating 
discussion on this topic.
This research has made use of the 
NASA Astrophysics Data System (ADS),
NIST Atomic Spectra Database, 
and the SIMBAD database, operated at CDS, Strasbourg, France.
A.F.\ acknowledges support through the W.J.~McDonald Fellowship of
the McDonald Observatory.
Funding for this project has also been generously provided by 
the Deutsche Forschungsgemeinschaft (contract KR~806/13-1),
the Helmholtz Gemeinschaft (grant VH-VI-061), and 
the U.~S.\ National Science Foundation
(grants AST~07-07447 to J.J.C.\ and AST~06-07708 to C.S.).


{\it Facilities:} 
\facility{HST (STIS)}
\facility{Keck:I (HIRES)}
\facility{VLT:Kueyen (UVES)}

\clearpage
\input{tab1}

\clearpage
\begin{landscape}
\input{tab2}

\clearpage
\end{landscape}

\clearpage
\begin{landscape}
\input{tab3}

\clearpage
\end{landscape}

\clearpage
\LongTables
\begin{landscape}
\input{tab4}

\clearpage
\end{landscape}

\clearpage
\vspace*{0.5in}
\input{tab5}

\vspace*{1.0in}
\input{tab6}

\clearpage
\vspace*{0.5in}
\input{tab7}

\vspace*{1.0in}
\input{tab8}

\clearpage
\begin{landscape}
\vspace*{2in}
\input{tab9}

\clearpage
\end{landscape}

\end{document}

%% file: tab1.tex
\begin{deluxetable}{lcccccccc}
\tablecaption{Adopted Atmospheric Parameters and Spectral Coverage
\label{atmtab}}
\tablewidth{0pt}
\tablehead{
\colhead{Star} &
\colhead{$T_{\rm eff}$} &
\colhead{log\,$g$} &
\colhead{$v_{t}$} &
\colhead{[Fe/H]} &
\colhead{Reference} &
\colhead{\textit{HST}} &
\colhead{Keck} &
\colhead{VLT} \\
\colhead{} &
\colhead{(K)} &
\colhead{} &
\colhead{(km\,s$^{-1}$)} &
\colhead{} &
\colhead{} &
\colhead{STIS} &
\colhead{HIRES} &
\colhead{UVES} }
\startdata
\bdcowan   & 5200 & 1.80 & 1.90 & $-$2.08 & 1 &  Y & Y & N \\
\cssneden  & 4800 & 1.50 & 1.95 & $-$3.10 & 2 &  Y & Y & N \\
\cshayek   & 5300 & 2.80 & 1.60 & $-$2.60 & 3 &  N & N & Y \\
HD~6268    & 4685 & 1.50 & 2.00 & $-$2.42 & 4 &  Y & Y & N \\
HD~74462   & 4700 & 2.00 & 1.90 & $-$1.52 & 5 &  N & Y & N \\
HD~108317  & 5234 & 2.68 & 2.00 & $-$2.18 & 5 &  N & Y & N \\
HD~115444  & 4720 & 1.75 & 2.00 & $-$2.90 & 5 &  Y & Y & N \\
HD~122563  & 4570 & 1.35 & 2.90 & $-$2.72 & 5 &  Y & Y & N \\
HD~122956  & 4510 & 1.55 & 1.60 & $-$1.95 & 5 &  Y & Y & N \\
HD~126587  & 4795 & 1.95 & 2.00 & $-$2.93 & 4 &  Y & Y & N \\
HD~175305  & 5040 & 2.85 & 2.00 & $-$1.48 & 4 &  Y & Y & N \\
HD~186478  & 4600 & 1.45 & 2.00 & $-$2.56 & 5 &  Y & Y & N \\
HD~204543  & 4672 & 1.49 & 2.00 & $-$1.87 & 5 &  N & Y & N \\
\hehayek   & 5060 & 2.30 & 1.60 & $-$2.97 & 3 &  N & N & Y \\
\enddata
\tablerefs{
(1)~\citealt{cowan02};
(2)~\citealt{sneden03};
(3)~\citealt{hayek09};
(4)~\citealt{cowan05};
(5)~\citealt{simmerer04}
}
\end{deluxetable}

%% file: tab2.tex
\begin{deluxetable}{lccccccccccc} 
\tablecaption{Abundances Dervied from Individual Transitions (I)
\label{indivtab1}}
\tablewidth{0pt}
\tablecolumns{12}
\tablehead{
\colhead{$\lambda$ (\AA)} &
\colhead{species} &
\colhead{E.P.\ (eV)} &
\colhead{log($gf$)} &
\colhead{Ref.} &
\colhead{\bdcowan} &
\colhead{\cssneden} &
\colhead{\cshayek} &
\colhead{HD~6268} &
\colhead{HD~74462} &
\colhead{HD~108317} & 
\colhead{HD~115444} }
\startdata
3794.77 & La~\textsc{ii} & 0.24 & $+$0.21 & 1    & $-$0.58$\pm$0.15 & $-$0.91$\pm$0.10 & $-$0.75$\pm$0.15 & $-$1.08$\pm$0.15 & \nodata          & $-$0.96$\pm$0.20 & $-$1.38$\pm$0.10 \\
3988.51 & La~\textsc{ii} & 0.40 & $+$0.21 & 1    & $-$0.57$\pm$0.10 & $-$0.88$\pm$0.10 & $-$0.80$\pm$0.15 & $-$1.08$\pm$0.10 & $-$0.24$\pm$0.10 & $-$1.04$\pm$0.15 & $-$1.48$\pm$0.10 \\
3995.74 & La~\textsc{ii} & 0.17 & $-$0.06 & 1    & $-$0.54$\pm$0.10 & $-$0.86$\pm$0.10 & $-$0.73$\pm$0.15 & $-$1.05$\pm$0.10 & $-$0.24$\pm$0.10 & $-$0.97$\pm$0.15 & $-$1.43$\pm$0.10 \\
4086.71 & La~\textsc{ii} & 0.00 & $-$0.07 & 1    & $-$0.52$\pm$0.10 & $-$0.83$\pm$0.10 & $-$0.70$\pm$0.10 & $-$0.97$\pm$0.10 & $-$0.30$\pm$0.20 & $-$0.96$\pm$0.20 & $-$1.36$\pm$0.10 \\
4123.22 & La~\textsc{ii} & 0.32 & $+$0.13 & 1    & $-$0.59$\pm$0.15 & $-$0.90$\pm$0.15 & $-$0.79$\pm$0.10 & $-$1.12$\pm$0.15 & $-$0.30$\pm$0.20 & $-$1.07$\pm$0.15 & $-$1.48$\pm$0.15 \\
3724.93 & Eu~\textsc{ii} & 0.00 & $-$0.09 & 2    & $-$0.75$\pm$0.20 & $-$0.94$\pm$0.20 & $-$0.98$\pm$0.20 & $-$1.31$\pm$0.15 & $-$0.49$\pm$0.25 & $-$1.30$\pm$0.15 & $-$1.61$\pm$0.15 \\
3819.67 & Eu~\textsc{ii} & 0.00 & $+$0.51 & 2    & $-$0.81$\pm$0.15 & $-$0.94$\pm$0.15 & $-$0.97$\pm$0.10 & $-$1.41$\pm$0.15 & \nodata          & $-$1.35$\pm$0.15 & $-$1.68$\pm$0.10 \\
3907.11 & Eu~\textsc{ii} & 0.21 & $+$0.17 & 2    & $-$0.81$\pm$0.10 & $-$0.96$\pm$0.10 & $-$0.95$\pm$0.10 & $-$1.39$\pm$0.10 & $-$0.53$\pm$0.25 & $-$1.31$\pm$0.10 & $-$1.64$\pm$0.10 \\
3930.50 & Eu~\textsc{ii} & 0.21 & $+$0.27 & 2    & $-$0.77$\pm$0.20 & $-$0.97$\pm$0.20 & $-$0.97$\pm$0.20 & $-$1.30$\pm$0.20 & \nodata          & $-$1.28$\pm$0.20 & $-$1.63$\pm$0.20 \\
3971.97 & Eu~\textsc{ii} & 0.21 & $+$0.27 & 2    & $-$0.81$\pm$0.20 & $-$0.99$\pm$0.20 & $-$0.97$\pm$0.20 & $-$1.40$\pm$0.20 & \nodata          & $-$1.36$\pm$0.20 & $-$1.64$\pm$0.20 \\
4129.72 & Eu~\textsc{ii} & 0.00 & $+$0.22 & 2    & $-$0.77$\pm$0.10 & $-$0.98$\pm$0.10 & $-$0.95$\pm$0.10 & $-$1.39$\pm$0.10 & $-$0.49$\pm$0.15 & $-$1.32$\pm$0.10 & $-$1.63$\pm$0.10 \\
4205.04 & Eu~\textsc{ii} & 0.00 & $+$0.21 & 2    & $-$0.77$\pm$0.15 & $-$0.97$\pm$0.20 & $-$0.97$\pm$0.15 & $-$1.36$\pm$0.15 & $-$0.52$\pm$0.20 & \nodata          & $-$1.62$\pm$0.10 \\
4435.58 & Eu~\textsc{ii} & 0.21 & $-$0.11 & 2    & $-$0.73$\pm$0.20 & \nodata          & $-$0.96$\pm$0.15 & \nodata          & $-$0.50$\pm$0.20 & \nodata          & $-$1.66$\pm$0.15 \\
3230.58 & Er~\textsc{ii} & 0.06 & $+$0.24 & 3    & $-$0.43$\pm$0.20 & $-$0.62$\pm$0.30 & \nodata          & $-$0.98$\pm$0.15 & $-$0.16$\pm$0.30 & $-$0.92$\pm$0.20 & $-$1.29$\pm$0.20 \\
3312.43 & Er~\textsc{ii} & 0.06 & $-$0.03 & 3    & $-$0.32$\pm$0.20 & $-$0.56$\pm$0.25 & \nodata          & $-$0.85$\pm$0.20 & \nodata          & $-$0.76$\pm$0.25 & $-$1.20$\pm$0.15 \\
3729.52 & Er~\textsc{ii} & 0.00 & $-$0.59 & 3    & $-$0.27$\pm$0.15 & $-$0.44$\pm$0.15 & $-$0.54$\pm$0.15 & $-$0.85$\pm$0.15 & $-$0.05$\pm$0.15 & $-$0.82$\pm$0.15 & $-$1.14$\pm$0.15 \\
3830.48 & Er~\textsc{ii} & 0.00 & $-$0.22 & 3    & $-$0.36$\pm$0.15 & $-$0.51$\pm$0.15 & $-$0.50$\pm$0.15 & $-$0.96$\pm$0.15 & $-$0.24$\pm$0.20 & $-$0.77$\pm$0.15 & $-$1.23$\pm$0.15 \\
3896.23 & Er~\textsc{ii} & 0.06 & $-$0.12 & 3    & $-$0.34$\pm$0.15 & $-$0.50$\pm$0.15 & $-$0.55$\pm$0.15 & $-$1.01$\pm$0.20 & $-$0.26$\pm$0.30 & $-$0.92$\pm$0.15 & $-$1.29$\pm$0.15 \\
3906.31 & Er~\textsc{ii} & 0.00 & $+$0.12 & 3    & $-$0.38$\pm$0.20 & $-$0.52$\pm$0.20 & $-$0.47$\pm$0.20 & \nodata          & \nodata          & $-$0.88$\pm$0.20 & $-$1.21$\pm$0.20 \\
3193.53 & Hf~\textsc{ii} & 0.38 & $-$0.89 & 4    & $-$0.85$\pm$0.25 & \nodata          & \nodata          & \nodata          & $-$0.29$\pm$0.30 & $< -$0.80        & \nodata          \\
3505.22 & Hf~\textsc{ii} & 1.04 & $-$0.14 & 4    & $-$0.75$\pm$0.20 & $-$1.07$\pm$0.30 & \nodata          & $-$1.09$\pm$0.25 & \nodata          & $-$1.00$\pm$0.20 & $-$1.52$\pm$0.30 \\
3918.09 & Hf~\textsc{ii} & 0.45 & $-$1.14 & 4    & \nodata          & $-$0.81$\pm$0.20 & $-$0.57$\pm$0.25 & $-$1.04$\pm$0.25 & \nodata          & \nodata          & \nodata          \\
4093.15 & Hf~\textsc{ii} & 0.45 & $-$1.15 & 4    & $-$0.68$\pm$0.15 & $-$0.89$\pm$0.15 & $-$0.81$\pm$0.30 & $-$1.24$\pm$0.20 & $-$0.39$\pm$0.10 & $-$1.00$\pm$0.30 & $-$1.47$\pm$0.30 \\
3513.65 & Ir~\textsc{i}  & 0.00 & $-$1.21 & 5    & $+$0.19$\pm$0.25 & $-$0.07$\pm$0.20 & $+$0.21$\pm$0.30 & $-$0.41$\pm$0.20 & $+$0.59$\pm$0.25 & $-$0.22$\pm$0.25 & $-$0.49$\pm$0.25 \\
3800.12 & Ir~\textsc{i}  & 0.00 & $-$1.44 & 5    & $+$0.08$\pm$0.20 & $-$0.14$\pm$0.20 & $+$0.06$\pm$0.30 & $-$0.57$\pm$0.20 & $+$0.37$\pm$0.15 & $-$0.27$\pm$0.20 & $-$0.86$\pm$0.25 \\
2833.03 & Pb~\textsc{i}  & 0.00 & $-$0.50 & 6, 7 & $< +$0.27        & $< -$0.15        & \nodata          & \nodata          & \nodata          & \nodata          & $< -$0.45        \\
3683.46 & Pb~\textsc{i}  & 0.97 & $-$0.54 & 6, 7 & $< +$0.42        & $< +$0.20        & $< +$0.35        & $< +$0.18        & $+$0.53$\pm$0.20 & $+$0.17$\pm$0.30 & $< +$0.05        \\
4057.81 & Pb~\textsc{i}  & 1.32 & $-$0.22 & 6, 7 & $< +$0.72        & $< +$0.35        & $< +$0.50        & $< +$0.08        & $< +$0.48        & $< +$0.37        & $< -$0.30        \\
3539.59 & Th~\textsc{ii} & 0.00 & $-$0.54 & 8    & $-$1.31$\pm$0.20 & $-$1.55$\pm$0.25 & \nodata          & $< -$1.58        & $-$1.17$\pm$0.30 & $< -$1.44        & $< -$1.86        \\
4019.13 & Th~\textsc{ii} & 0.00 & $-$0.23 & 8    & $-$1.27$\pm$0.15 & $-$1.68$\pm$0.20 & $-$1.46$\pm$0.25 & $-$1.78$\pm$0.15 & $-$0.88$\pm$0.20 & $-$1.84$\pm$0.20 & $-$2.08$\pm$0.15 \\
4086.52 & Th~\textsc{ii} & 0.00 & $-$0.93 & 8    & $-$1.14$\pm$0.30 & $-$1.48$\pm$0.30 & $< -$1.06        & $< -$1.33        & $-$0.90$\pm$0.20 & $< -$1.19        & $< -$1.71        \\
4094.75 & Th~\textsc{ii} & 0.00 & $-$0.88 & 8    & $-$1.19$\pm$0.30 & $-$1.60$\pm$0.30 & $< -$1.16        & $< -$1.43        & $< -$0.88        & $< -$1.04        & $< -$1.81        \\
\enddata
\tablerefs{
(1)~\citealt{lawler01a};
(2)~\citealt{lawler01b};
(3)~\citealt{lawler08}; 
(4)~\citealt{lawler07};
(5)~\citealt{ivarsson03}, with updates as noted in the Appendix of \citealt{cowan05};
(6)~\citealt{wood68};
(7)~\citealt{biemont00};
(8)~\citealt{nilsson02}
}
\end{deluxetable}

%% file: tab3.tex
\begin{deluxetable}{lccccccccccc} 
\tablecaption{Abundances Dervied from Individual Transitions (II)
\label{indivtab2}}
\tablewidth{0pt}
\tablecolumns{12}
\tablehead{
\colhead{$\lambda$ (\AA)} &
\colhead{species} &
\colhead{E.P.\ (eV)} &
\colhead{log($gf$)} &
\colhead{Ref.} &
\colhead{HD~122563} & 
\colhead{HD~122956} &
\colhead{HD~126587} &
\colhead{HD~175305} &
\colhead{HD~186478} &
\colhead{HD~204543} & 
\colhead{\hehayek} }
\startdata
3794.77 & La~\textsc{ii} & 0.24 & $+$0.21 & 1    & \nodata          & $-$0.78$\pm$0.25 & $-$1.75$\pm$0.15 & $-$0.16$\pm$0.20 & $-$1.34$\pm$0.20 & \nodata          & $-$0.75$\pm$0.20 \\
3988.51 & La~\textsc{ii} & 0.40 & $+$0.21 & 1    & $-$2.50$\pm$0.30 & $-$0.63$\pm$0.10 & $-$1.83$\pm$0.25 & $-$0.14$\pm$0.10 & $-$1.33$\pm$0.10 & $-$0.63$\pm$0.10 & $-$0.80$\pm$0.20 \\
3995.74 & La~\textsc{ii} & 0.17 & $-$0.06 & 1    & $-$2.32$\pm$0.30 & $-$0.63$\pm$0.10 & $-$1.82$\pm$0.15 & $-$0.12$\pm$0.10 & $-$1.34$\pm$0.10 & $-$0.62$\pm$0.10 & $-$0.73$\pm$0.15 \\
4086.71 & La~\textsc{ii} & 0.00 & $-$0.07 & 1    & $-$2.34$\pm$0.20 & $-$0.66$\pm$0.10 & $-$1.59$\pm$0.10 & $-$0.17$\pm$0.10 & $-$1.28$\pm$0.10 & $-$0.63$\pm$0.10 & $-$0.72$\pm$0.10 \\
4123.22 & La~\textsc{ii} & 0.32 & $+$0.13 & 1    & $-$2.48$\pm$0.25 & $-$0.58$\pm$0.20 & $-$1.81$\pm$0.20 & $-$0.13$\pm$0.15 & $-$1.35$\pm$0.15 & $-$0.65$\pm$0.15 & $-$0.88$\pm$0.20 \\
3724.93 & Eu~\textsc{ii} & 0.00 & $-$0.09 & 2    & $-$2.77$\pm$0.25 & $-$0.85$\pm$0.25 & $-$1.90$\pm$0.20 & $-$0.31$\pm$0.15 & $-$1.49$\pm$0.15 & $-$1.03$\pm$0.15 & $-$0.97$\pm$0.20 \\
3819.67 & Eu~\textsc{ii} & 0.00 & $+$0.51 & 2    & $-$2.91$\pm$0.30 & \nodata          & $-$2.04$\pm$0.15 & \nodata          & $-$1.56$\pm$0.10 & \nodata          & $-$1.00$\pm$0.10 \\
3907.11 & Eu~\textsc{ii} & 0.21 & $+$0.17 & 2    & $-$2.74$\pm$0.25 & $-$0.98$\pm$0.15 & $-$1.97$\pm$0.15 & $-$0.38$\pm$0.20 & $-$1.55$\pm$0.10 & $-$1.14$\pm$0.15 & $-$0.99$\pm$0.10 \\
3930.50 & Eu~\textsc{ii} & 0.21 & $+$0.27 & 2    & $< -$2.01        & \nodata          & $-$1.96$\pm$0.20 & \nodata          & \nodata          & \nodata          & $-$1.03$\pm$0.20 \\
3971.97 & Eu~\textsc{ii} & 0.21 & $+$0.27 & 2    & $< -$2.51        & $-$1.03$\pm$0.20 & $-$1.95$\pm$0.20 & \nodata          & $-$1.53$\pm$0.20 & \nodata          & $-$1.04$\pm$0.20 \\
4129.72 & Eu~\textsc{ii} & 0.00 & $+$0.22 & 2    & $-$2.71$\pm$0.20 & $-$0.93$\pm$0.15 & $-$1.95$\pm$0.10 & $-$0.37$\pm$0.10 & $-$1.51$\pm$0.10 & $-$1.04$\pm$0.15 & $-$0.97$\pm$0.10 \\
4205.04 & Eu~\textsc{ii} & 0.00 & $+$0.21 & 2    & $-$2.68$\pm$0.30 & $-$0.93$\pm$0.20 & $-$2.02$\pm$0.20 & $-$0.38$\pm$0.25 & $-$1.53$\pm$0.15 & $-$1.05$\pm$0.15 & $-$0.95$\pm$0.15 \\
4435.58 & Eu~\textsc{ii} & 0.21 & $-$0.11 & 2    & $< -$2.21        & $-$0.88$\pm$0.20 & $< -$1.82        & $-$0.40$\pm$0.25 & $-$1.55$\pm$0.20 & $-$0.99$\pm$0.15 & $-$0.91$\pm$0.20 \\
3230.58 & Er~\textsc{ii} & 0.06 & $+$0.24 & 3    & $-$2.38$\pm$0.30 & $-$0.31$\pm$0.20 & $-$1.47$\pm$0.25 & $-$0.06$\pm$0.20 & $-$1.10$\pm$0.20 & $-$0.68$\pm$0.25 & \nodata          \\
3312.43 & Er~\textsc{ii} & 0.06 & $-$0.03 & 3    & $< -$2.11        & $-$0.24$\pm$0.20 & $-$1.47$\pm$0.30 & $-$0.02$\pm$0.30 & $-$1.06$\pm$0.25 & $-$0.48$\pm$0.25 & \nodata          \\
3729.52 & Er~\textsc{ii} & 0.00 & $-$0.59 & 3    & $< -$1.76        & $-$0.45$\pm$0.15 & $-$1.43$\pm$0.20 & $+$0.03$\pm$0.15 & $-$1.03$\pm$0.15 & $-$0.63$\pm$0.15 & $-$0.49$\pm$0.15 \\
3830.48 & Er~\textsc{ii} & 0.00 & $-$0.22 & 3    & $-$2.23$\pm$0.30 & $-$0.46$\pm$0.15 & $-$1.46$\pm$0.15 & $-$0.03$\pm$0.15 & $-$1.13$\pm$0.15 & $-$0.75$\pm$0.20 & $-$0.52$\pm$0.15 \\
3896.23 & Er~\textsc{ii} & 0.06 & $-$0.12 & 3    & $< -$2.16        & $-$0.49$\pm$0.25 & $-$1.56$\pm$0.20 & $+$0.00$\pm$0.20 & $-$1.16$\pm$0.20 & $-$0.58$\pm$0.25 & $-$0.50$\pm$0.15 \\
3906.31 & Er~\textsc{ii} & 0.00 & $+$0.12 & 3    & \nodata          & \nodata          & $-$1.46$\pm$0.20 & \nodata          & $-$1.03$\pm$0.25 & \nodata          & $-$0.45$\pm$0.20 \\
3193.53 & Hf~\textsc{ii} & 0.38 & $-$0.89 & 4    & \nodata          & \nodata          & \nodata          & $-$0.25$\pm$0.30 & \nodata          & $-$0.89$\pm$0.30 & \nodata          \\
3505.22 & Hf~\textsc{ii} & 1.04 & $-$0.14 & 4    & $< -$1.64        & $-$0.69$\pm$0.25 & $< -$1.40        & $-$0.22$\pm$0.20 & $-$1.23$\pm$0.25 & $-$0.77$\pm$0.20 & \nodata          \\
3918.09 & Hf~\textsc{ii} & 0.45 & $-$1.14 & 4    & \nodata          & $-$0.48$\pm$0.15 & $< -$1.05        & \nodata          & \nodata          & \nodata          & $< -$0.54        \\
4093.15 & Hf~\textsc{ii} & 0.45 & $-$1.15 & 4    & $< -$1.89        & $-$0.77$\pm$0.15 & $-$1.65$\pm$0.30 & $-$0.05$\pm$0.15 & $-$1.40$\pm$0.15 & $-$0.83$\pm$0.15 & $< -$0.39        \\
3513.65 & Ir~\textsc{i}  & 0.00 & $-$1.21 & 5    & $< -$0.51        & $-$0.08$\pm$0.20 & $< -$0.47        & $+$0.40$\pm$0.20 & $-$0.62$\pm$0.25 & $-$0.13$\pm$0.25 & \nodata          \\
3800.12 & Ir~\textsc{i}  & 0.00 & $-$1.44 & 5    & $< -$1.16        & $-$0.09$\pm$0.20 & $-$1.05$\pm$0.25 & $+$0.41$\pm$0.20 & $-$0.79$\pm$0.20 & $-$0.27$\pm$0.20 & $-$0.19$\pm$0.25 \\
2833.03 & Pb~\textsc{i}  & 0.00 & $-$0.50 & 6, 7 & $< -$0.42        & $-$0.15$\pm$0.20 & $< -$0.38        & $+$0.07$\pm$0.30 & \nodata          & \nodata          & \nodata          \\
3683.46 & Pb~\textsc{i}  & 0.97 & $-$0.54 & 6, 7 & $< -$0.02        & $+$0.00$\pm$0.20 & $< +$0.07        & $+$0.22$\pm$0.30 & $< -$0.01        & $+$0.05$\pm$0.25 & $< +$0.63        \\
4057.81 & Pb~\textsc{i}  & 1.32 & $-$0.22 & 6, 7 & $< -$0.27        & $< +$0.05        & $< -$0.28        & $< +$0.42        & $< -$0.26        & $< +$0.13        & $< +$0.53        \\
3539.59 & Th~\textsc{ii} & 0.00 & $-$0.54 & 8    & $< -$1.88        & $-$1.76$\pm$0.30 & $< -$1.84        & $-$0.84$\pm$0.30 & $< -$1.92        & $-$1.96$\pm$0.30 & $< -$0.48        \\
4019.13 & Th~\textsc{ii} & 0.00 & $-$0.23 & 8    & $< -$2.43        & $-$1.38$\pm$0.20 & $< -$2.39        & $-$0.74$\pm$0.20 & $-$2.19$\pm$0.25 & $-$1.53$\pm$0.25 & $-$1.25$\pm$0.25 \\
4086.52 & Th~\textsc{ii} & 0.00 & $-$0.93 & 8    & $< -$2.08        & $< -$1.66        & $< -$1.79        & $< -$0.64        & $< -$2.02        & $-$1.71$\pm$0.30 & $-$1.03$\pm$0.40 \\
4094.75 & Th~\textsc{ii} & 0.00 & $-$0.88 & 8    & $< -$1.93        & $< -$1.41        & $< -$1.79        & $-$0.71$\pm$0.30 & $< -$1.92        & $-$1.60$\pm$0.30 & $< -$0.88        \\
\enddata
\tablerefs{
(1)~\citealt{lawler01a};
(2)~\citealt{lawler01b};
(3)~\citealt{lawler08}; 
(4)~\citealt{lawler07};
(5)~\citealt{ivarsson03}, with updates as noted in the Appendix of \citealt{cowan05};
(6)~\citealt{wood68};
(7)~\citealt{biemont00};
(8)~\citealt{nilsson02}
}
\end{deluxetable}

%% file: tab4.tex
\begin{deluxetable}{lccccccccccc}
\tablecaption{Final Elemental Abundances
\label{abundtab}}
\tablewidth{0pt}
\tablehead{
\colhead{Star} &
\colhead{[Fe/H]} &
\colhead{Ref.} &
\colhead{\eps{La}} &
\colhead{\eps{Eu}} &
\colhead{\eps{Er}} &
\colhead{\eps{Hf}} &
\colhead{\eps{Ir}} &
\colhead{\eps{Pb}} &
\colhead{\eps{Th}} &
\colhead{Ref.} &
\colhead{$r$-only?\tablenotemark{a}}}
\startdata
BD$-$18~5550     & $-$3.05 & 1  &  $-$2.52$\pm$0.10  & $-$2.81$\pm$0.20  & $-$2.37$\pm$0.10 & $< -$0.86        & \nodata          & \nodata             & $< -$3.02         & 1, 2        &   \\
BD$+$01~2916     & $-$1.93 & 3  &  $-$0.87$\pm$0.12  & $-$1.22$\pm$0.14  & \nodata          & \nodata          & \nodata          & $-$0.20$\pm$0.19    & \nodata           & 3	          &   \\
BD$+$04~2621     & $-$2.52 & 1  &  $-$2.29$\pm$0.11  & $-$2.63$\pm$0.20  & $-$2.24$\pm$0.10 & \nodata          & \nodata          & \nodata	        & $< -$3.04         & 1, 2        &   \\
BD$+$06~0648     & $-$2.14 & 3  &  $-$0.95$\pm$0.12  & $-$1.50$\pm$0.16  & \nodata          & \nodata          & \nodata          & $< +$0.00	        & \nodata           & 3	          &   \\
BD$+$08~2856     & $-$2.12 & 1  &  $-$1.03$\pm$0.04  & $-$1.16$\pm$0.04  & $-$0.95$\pm$0.09 & $< +$0.20        & \nodata          & \nodata	        & $-$1.70$\pm$0.10  & 1, 2        & Y \\
BD$+$17~3248     & $-$2.08 & 4  &  $-$0.55$\pm$0.05  & $-$0.78$\pm$0.05  & $-$0.34$\pm$0.07 & $-$0.73$\pm$0.11 & $+$0.12$\pm$0.16 & $< +$0.27	        & $-$1.26$\pm$0.10  & 5	          & Y \\
BD$+$30~2611     & $-$1.46 & 3  &  $-$0.27$\pm$0.12  & $-$0.49$\pm$0.14  & \nodata          & \nodata          & \nodata          & $+$0.43$\pm$0.19    & \nodata           & 3	          & Y \\
CS~22892--052    & $-$3.10 & 6  &  $-$0.87$\pm$0.05  & $-$0.96$\pm$0.05  & $-$0.50$\pm$0.07 & $-$0.93$\pm$0.13 & $-$0.10$\pm$0.14 & $< -$0.15	        & $-$1.60$\pm$0.13  & 5	          & Y \\
CS~29491--069    & $-$2.60 & 7  &  $-$0.75$\pm$0.05  & $-$0.96$\pm$0.05  & $-$0.52$\pm$0.08 & $-$0.67$\pm$0.19 & $+$0.13$\pm$0.21 & $< +$0.35	        & $-$1.46$\pm$0.25  & 5	          & Y \\
CS~29497--004    & $-$2.63 & 8  &  $-$0.38$\pm$0.15  & $-$0.45$\pm$0.20  & \nodata          & \nodata          & \nodata          & \nodata	        & $-$0.96$\pm$0.15  & 8, 9        & Y \\
CS~30306--132    & $-$2.50 & 10 &  $-$0.78$\pm$0.06  & $-$1.02$\pm$0.05  & $-$0.62$\pm$0.15 & \nodata          & \nodata          & $< +$0.50	        & $-$1.16$\pm$0.15  & 10          & Y \\
CS~31078--018    & $-$2.84 & 11 &  $-$1.00$\pm$0.22  & $-$1.17$\pm$0.17  & $-$0.99$\pm$0.15 & \nodata          & \nodata          & $< +$0.25	        & $-$1.35$\pm$0.25  & 12          & Y \\
CS~31082--001    & $-$2.90 & 12 &  $-$0.62$\pm$0.04  & $-$0.72$\pm$0.03  & $-$0.30$\pm$0.04 & $-$0.72$\pm$0.04 & $+$0.22$\pm$0.20 & $-$0.55$\pm$0.15    & $-$0.98$\pm$0.05  & 12, 13, 14  & Y \\
HD~3008          & $-$1.98 & 3  &  $-$1.02$\pm$0.14  & $-$1.30$\pm$0.16  & \nodata          & \nodata          & \nodata          & $-$0.35$\pm$0.22    & \nodata  	    & 3	          &   \\
HD~6268          & $-$2.42 & 15 &  $-$1.05$\pm$0.05  & $-$1.37$\pm$0.05  & $-$0.93$\pm$0.07 & $-$1.14$\pm$0.13 & $-$0.49$\pm$0.14 & $< +$0.08	        & $-$1.78$\pm$0.15  & 5	          &   \\
HD~29574         & $-$1.86 & 3  &  $-$0.63$\pm$0.12  & $-$0.90$\pm$0.14  & \nodata          & \nodata          & \nodata          & $-$0.15$\pm$0.19    & \nodata  	    & 3	          &   \\
HD~74462         & $-$1.52 & 16 &  $-$0.25$\pm$0.06  & $-$0.50$\pm$0.09  & $-$0.14$\pm$0.10 & $-$0.38$\pm$0.09 & $+$0.43$\pm$0.13 & $+$0.53$\pm$0.20    & $-$0.94$\pm$0.13  & 5	          &   \\
HD~108317        & $-$2.18 & 16 &  $-$1.01$\pm$0.07  & $-$1.32$\pm$0.05  & $-$0.85$\pm$0.07 & $-$1.00$\pm$0.17 & $-$0.25$\pm$0.16 & $+$0.17$\pm$0.30    & $-$1.84$\pm$0.20  & 5	          &   \\
HD~108577        & $-$2.38 & 1  &  $-$1.24$\pm$0.09  & $-$1.48$\pm$0.12  & $-$1.23$\pm$0.11 & \nodata          & \nodata          & \nodata	        & $-$2.03$\pm$0.14  & 1, 2        & Y \\
HD~115444        & $-$2.90 & 16 &  $-$1.42$\pm$0.05  & $-$1.64$\pm$0.04  & $-$1.22$\pm$0.07 & $-$1.50$\pm$0.21 & $-$0.68$\pm$0.18 & $< -$0.45	        & $-$2.08$\pm$0.15  & 5	          & Y \\
HD~122563        & $-$2.72 & 16 &  $-$2.40$\pm$0.13  & $-$2.75$\pm$0.11  & $-$2.30$\pm$0.21 & $< -1.89$        & $< -1.16$        & $< -$0.42	        & $< -$2.43	    & 5	          &   \\
HD~122956        & $-$1.95 & 16 &  $-$0.64$\pm$0.05  & $-$0.94$\pm$0.07  & $-$0.40$\pm$0.08 & $-$0.83$\pm$0.08 & $-$0.08$\pm$0.14 & $-$0.08$\pm$0.14    & $-$1.50$\pm$0.17  & 5	          &   \\
HD~126587        & $-$2.93 & 15 &  $-$1.75$\pm$0.07  & $-$1.97$\pm$0.06  & $-$1.47$\pm$0.08 & $-$1.65$\pm$0.30 & $-$1.05$\pm$0.25 & $< -$0.38           & $< -$2.39	    & 5	          & Y \\
HD~128279        & $-$2.40 & 1  &  $-$1.51$\pm$0.16  & $-$1.80$\pm$0.20  & $-$1.41$\pm$0.10 & \nodata          & \nodata          & \nodata	        & $< -$1.88	    & 1, 2        &   \\
HD~141531        & $-$1.66 & 3  &  $-$0.55$\pm$0.12  & $-$0.87$\pm$0.14  & \nodata          & \nodata          & \nodata          & $+$0.05$\pm$0.19    & \nodata  	    & 3	          &   \\
HD~175305        & $-$1.48 & 15 &  $-$0.14$\pm$0.05  & $-$0.36$\pm$0.07  & $-$0.01$\pm$0.08 & $-$0.13$\pm$0.11 & $+$0.40$\pm$0.14 & $+$0.25$\pm$0.21    & $-$0.76$\pm$0.15  & 5	          & Y \\
HD~186478        & $-$2.56 & 16 &  $-$1.32$\pm$0.05  & $-$1.53$\pm$0.05  & $-$1.09$\pm$0.08 & $-$1.35$\pm$0.13 & $-$0.72$\pm$0.16 & $< -$0.26	        & $-$2.19$\pm$0.25  & 5	          & Y \\
HD~204543        & $-$1.87 & 16 &  $-$0.63$\pm$0.05  & $-$1.05$\pm$0.07  & $-$0.64$\pm$0.09 & $-$0.82$\pm$0.11 & $-$0.21$\pm$0.16 & $+$0.05$\pm$0.25    & $-$1.68$\pm$0.14  & 5	          &   \\
HD~206739        & $-$1.64 & 3  &  $-$0.41$\pm$0.12  & $-$0.72$\pm$0.18  & \nodata          & \nodata          & \nodata          & $+$0.20$\pm$0.26    & \nodata  	    & 3	          &   \\
HD~214925        & $-$2.08 & 3  &  $-$0.86$\pm$0.12  & $-$1.09$\pm$0.20  & \nodata          & \nodata          & \nodata          & $-$0.50$\pm$0.22    & \nodata  	    & 3	          & Y \\
HD~216143        & $-$2.32 & 3  &  $-$1.21$\pm$0.12  & $-$1.24$\pm$0.15  & \nodata          & \nodata          & \nodata          & $< -$0.10	        & \nodata  	    & 3	          & Y \\
HD~220838        & $-$1.80 & 3  &  $-$0.76$\pm$0.12  & $-$0.93$\pm$0.16  & \nodata          & \nodata          & \nodata          & $+$0.05$\pm$0.19    & \nodata  	    & 3	          & Y \\
HD~221170        & $-$2.16 & 17 &  $-$0.73$\pm$0.06  & $-$0.86$\pm$0.07  & $-$0.47$\pm$0.08 & $-$0.84$\pm$0.11 & $+$0.02$\pm$0.13 & $-$0.09$\pm$0.21    & $-$1.46$\pm$0.05  & 17          & Y \\
HD~235766        & $-$1.93 & 3  &  $-$0.60$\pm$0.12  & $-$0.86$\pm$0.14  & \nodata          & \nodata          & \nodata          & $+$0.10$\pm$0.26    & \nodata	    & 3	          &   \\
HE~1219--0312    & $-$2.97 & 7  &  $-$0.75$\pm$0.07  & $-$0.98$\pm$0.05  & $-$0.49$\pm$0.05 & $< -0.89$        & $-$0.19$\pm$0.25 & $< +$0.53	        & $-$1.19$\pm$0.21  & 5	          & Y \\
HE~1523--0901    & $-$2.95 & 18 &  $-$0.63$\pm$0.06  & $-$0.62$\pm$0.05  & $-$0.42$\pm$0.17 & $-$0.73$\pm$0.20 & $+$0.24$\pm$0.05 & $< -$0.20	        & $-$1.20$\pm$0.05  & 18          & Y \\
M5~IV-81	 & $-$1.28 & 19 &  $+$0.11$\pm$0.05  & $-$0.31$\pm$0.05  & \nodata          & $-$0.12$\pm$0.15 & \nodata          & $+$0.35$\pm$0.14    & $-$0.58$\pm$0.15  & 19, 20	  &   \\
M5~IV-82	 & $-$1.33 & 19 &  $+$0.11$\pm$0.05  & $-$0.23$\pm$0.05  & \nodata          & $-$0.22$\pm$0.15 & \nodata          & $+$0.25$\pm$0.14    & $-$0.68$\pm$0.15  & 19, 20	  &   \\
M13~L598	 & $-$1.56 & 21 &  $-$0.34$\pm$0.07  & $-$0.58$\pm$0.08  & \nodata          & \nodata          & \nodata          & $+$0.09$\pm$0.13    & \nodata  	    & 20, 21 	  & Y \\
M13~L629	 & $-$1.63 & 21 &  $-$0.35$\pm$0.07  & $-$0.61$\pm$0.08  & \nodata          & \nodata          & \nodata          & $+$0.12$\pm$0.13    & \nodata  	    & 20, 21	  &   \\
M13~L70	 	 & $-$1.59 & 21 &  $-$0.23$\pm$0.07  & $-$0.58$\pm$0.08  & \nodata          & \nodata          & \nodata          & $+$0.09$\pm$0.13    & \nodata  	    & 20, 21	  &   \\
M13~L973	 & $-$1.61 & 21 &  $-$0.27$\pm$0.07  & $-$0.51$\pm$0.08  & \nodata          & \nodata          & \nodata          & $-$0.01$\pm$0.13    & \nodata           & 20, 21	  & Y \\
M15~K341         & $-$2.32 & 22 &  $-$0.73$\pm$0.08  & $-$0.88$\pm$0.09  & \nodata          & \nodata          & \nodata          & \nodata             & $-$1.51$\pm$0.10  & 22          & Y \\
M15~K462         & $-$2.25 & 22 &  $-$0.47$\pm$0.08  & $-$0.61$\pm$0.09  & \nodata          & \nodata          & \nodata          & \nodata             & $-$1.30$\pm$0.10  & 22          & Y \\
M15~K583         & $-$2.34 & 22 &  $-$1.19$\pm$0.08  & $-$1.24$\pm$0.09  & \nodata          & \nodata          & \nodata          & \nodata             & $-$1.70$\pm$0.10  & 22          & Y \\
M92~VII-18       & $-$2.29 & 1  &  $-$1.29$\pm$0.07  & $-$1.45$\pm$0.07  & $-$1.14$\pm$0.18 & \nodata          & \nodata          & \nodata	        & $-$2.01$\pm$0.07  & 1, 2        & Y \\
UMi~COS82	 & $-$1.42 & 23 &  $+$0.52$\pm$0.16  & $+$0.34$\pm$0.11  & $+$0.73$\pm$0.11 & \nodata          & \nodata          & \nodata	        & $-$0.25$\pm$0.15  & 23          & Y \\
\enddata
\tablenotetext{a}{Defined by \eps{La/Eu}~$<+0.25$}
\tablerefs{
(1)~\citealt{johnson02a};
(2)~\citealt{johnson01};
(3)~\citealt{aoki08b};
(4)~\citealt{cowan02};
(5)~this study;
(6)~\citealt{sneden03};
(7)~\citealt{hayek09};
(8)~\citealt{christlieb04};
(9)~\citealt{jonsell06};
(10)~\citealt{honda04};
(11)~\citealt{lai08};
(12)~\citealt{hill02};
(13)~\citealt{plez04};
(14)~\citealt{sneden09};
(15)~\citealt{cowan05};
(16)~\citealt{simmerer04};
(17)~\citealt{ivans06};
(18)~\citealt{frebel07};
(19)~\citealt{yong08a};
(20)~\citealt{yong08b};
(21)~\citealt{yong06};
(22)~\citealt{sneden00};
(23)~\citealt{aoki07}
}
\end{deluxetable}

%% file: tab5.tex
\begin{deluxetable}{lcc} 
\tablecaption{$^{12}$C/$^{13}$C Ratios and Luminosities
\label{cratiotab}}
\tablewidth{0pt}
\tablecolumns{3}
\tablehead{
\colhead{Star} &
\colhead{$^{12}$C/$^{13}$C} &
\colhead{log($L/L_{\odot}$)} } 
\startdata
\bdcowan   &  6$\pm$3 & 2.4 \\
\cssneden  & 13$\pm$3 & 2.5 \\
\cshayek   & $>$30    & 1.4 \\
HD~6268    &  6$\pm$2 & 2.5 \\
HD~74462   & 10$\pm$3 & 2.0 \\
HD~108317  & 15$\pm$5 & 1.5 \\
HD~115444  &  6$\pm$2 & 2.2 \\
HD~122563  &  4$\pm$1 & 2.6 \\
HD~122956  &  7$\pm$2 & 2.4 \\ 
HD~126587  & 10$\pm$3 & 2.1 \\
HD~175305  & $>$25    & 1.3 \\
HD~186478  &  5$\pm$2 & 2.5 \\
HD~204543  &  5$\pm$2 & 2.5 \\
\hehayek   & $>$15    & 1.8 \\
\enddata
\end{deluxetable}

%% file: tab6.tex
\begin{deluxetable}{llccc}
\tablecaption{Pb, Th, and $^{12}$C/$^{13}$C Comparisons to Previous Studies
\label{comparetab}}
\tablewidth{0pt}
\tablehead{
\colhead{Star} &
\colhead{Reference} &
\colhead{\eps{Pb}} &
\colhead{\eps{Th}} &
\colhead{$^{12}$C/$^{13}$C}}
\startdata
\bdcowan      & this study          & $< +$0.27        & $-$1.26$\pm$0.10 &  6$\pm$3 \\
              & \citet{cowan02}     & $< +$0.3         & $-$1.22$\pm$0.10 & \nodata  \\
\hline
\cssneden     & this study          & $< -$0.15        & $-$1.60$\pm$0.13 & 13$\pm$3 \\
              & \citet{sneden03}    & $< -$0.2         & $-$1.59$\pm$0.10 & 15$\pm$2 \\
              & \citet{honda04}     & \nodata          & $-$1.46$\pm$0.15 & 20       \\
\hline
\cshayek      & this study          & $< +$0.35        & $-$1.46$\pm$0.25 & $>$30    \\
              & \citet{hayek09}     & \nodata          & $-$1.43$\pm$0.22 & $\gtrsim$90 \\
\hline
HD~6268       & this study          & $< +$0.08        & $-$1.78$\pm$0.15 & 6$\pm$2  \\
              & \citet{honda04}     & \nodata          & $-$1.97$\pm$0.10 & 4        \\
\hline
HD~74462      & this study          & $+$0.53$\pm$0.20 & $-$0.94$\pm$0.13 & 10$\pm$3 \\
              & \citet{aoki08b}     & $+$0.35$\pm$0.26 & \nodata          & \nodata  \\
\hline
HD~115444     & this study          & $< -$0.45        & $-$2.08$\pm$0.15 & 6$\pm$2  \\
              & \citet{westin00}    & \nodata          & $-$2.27$\pm$0.11 & 6$\pm$1.5 \\
              & \citet{johnson01}   & \nodata          & $-$2.40$\pm$0.09 & 6        \\
              & \citet{honda04}     & \nodata          & $-$2.01$\pm$0.15 & 7        \\
\hline
HD~186478     & this study          & $< -$0.26        & $-$2.19$\pm$0.25 & 5$\pm$2  \\
              & \citet{johnson01}   & \nodata          & $-$2.30$\pm$0.11 & 6        \\
              & \citet{honda04}     & \nodata          & $-$1.89$\pm$0.15 & 6        \\
\hline
HD~204543     & this study          & $+$0.05$\pm$0.25 & $-$1.68$\pm$0.14 & 5$\pm$2  \\
              & \citet{aoki08b}     & $+$0.00$\pm$0.22 & \nodata          & \nodata  \\
\hline
\hehayek      & this study          & $< +$0.53        & $-$1.19$\pm$0.21 & $>$15    \\
              & \citet{hayek09}     & \nodata          & $-$1.29$\pm$0.14 & $\gtrsim$90 \\
\enddata
\end{deluxetable}

%% file: tab7.tex
\begin{deluxetable}{ccccccccccccc} 
\tablecaption{Observed Present-Day \rpro\ Mean Ratios
\label{means}}
\tablewidth{0pt}
\tablecolumns{13}
\tablehead{
\colhead{} &
\colhead{S.S.\ $r$-only} &
\multicolumn{3}{c}{``standard'' stars with} &
\colhead{} &
\multicolumn{3}{c}{four ``standard''} &
\colhead{} &
\multicolumn{3}{c}{four stars with an} \\
\colhead{} &
\colhead{predictions\tablenotemark{a}} &
\multicolumn{3}{c}{\eps{La/Eu}~$< +0.25$} &
\colhead{} &
\multicolumn{3}{c}{$r$-only stars\tablenotemark{b}} &
\colhead{} &
\multicolumn{3}{c}{``actinide boost''\tablenotemark{c}} \\
\cline{3-5} \cline{7-9} \cline{11-13}
\colhead{Ratio} &
\colhead{($\log \epsilon$)} &
\colhead{$\langle \log \epsilon \rangle$} &
\colhead{$\sigma_{\mu}$} &
\colhead{No.} &
\colhead{} &
\colhead{$\langle \log \epsilon \rangle$} &
\colhead{$\sigma_{\mu}$} &
\colhead{No.} &
\colhead{} &
\colhead{$\langle \log \epsilon \rangle$} &
\colhead{$\sigma_{\mu}$} &
\colhead{No.} }
\startdata
La/Eu   & $+$0.179 & \multicolumn{3}{c}{\ldots} & & $+$0.18 & 0.03 & 4 & & $+$0.16 & 0.04 & 4 \\ 
Er/Eu   & $+$0.364 & $+$0.40 & 0.03 & 13        & & $+$0.44 & 0.03 & 4 & & $+$0.43 & 0.08 & 4 \\
Hf/Eu   & $-$0.075 & $+$0.10 & 0.04 &  9        & & $+$0.05 & 0.04 & 4 & & $+$0.00 & 0.06 & 1 \\
Ir/Eu   & $+$0.850 & $+$0.88 & 0.04 &  9        & & $+$0.89 & 0.05 & 4 & & $+$0.88 & 0.09 & 2 \\
Pb/Eu   & \nodata  & $+$0.68 & 0.07 &  7        & & $+$0.77 & 0.22 & 1 & & $+$0.17 & 0.15 & 1 \\
Th/Eu   & \nodata  & $-$0.56 & 0.03 & 16        & & $-$0.55 & 0.05 & 4 & & $-$0.24 & 0.03 & 4 \\
\enddata
\tablenotetext{a}{\citet{lodders03} S.S.\ meteoritic and \citet{sneden08},
with updates from Gallino}
\tablenotetext{b}{\bdcowan, \cssneden, \mbox{HD~115444}, \mbox{HD~221170}}
\tablenotetext{c}{\cshonda, \cslai, \cshill, \hehayek}
\end{deluxetable}

%% file: tab8.tex
\begin{deluxetable}{lcccccccc} 
\tablecaption{Comparison of Predicted and Calculated \rpro\ Abundances for Pb, Th, and U
\label{pbtab}}
\tablewidth{0pt}
\tablecolumns{9}
\tablehead{
\colhead{} &
\colhead{$^{206}$Pb} &
\colhead{$^{207}$Pb} &
\colhead{$^{208}$Pb} &
\colhead{$\sum$ Pb} &
\colhead{$^{232}$Th} &
\colhead{$^{235}$U} &
\colhead{$^{238}$U} &
\colhead{$\sum$ Th, U} }
\startdata
S.S.\ total \citep{lodders03}                & 0.601  & 0.665  & 1.903  & 3.169  & 0.0440 & 0.0059 & 0.0187 & 0.0686 \\
\hline
calculated \rpro:                            &        &        &        &        &        &        &        &        \\
\hspace*{3ex} ETFSI-Q:                       &        &        &        &        &        &        &        &        \\
\hspace*{6ex} Direct (isobaric) production   & 0.0209 & 0.0178 & 0.0283 & 0.0670 & 0.0184 & 0.0091 & 0.0076 & 0.0351 \\
\hspace*{6ex} Direct $+$ indirect production & 0.1439 & 0.1068 & 0.1242 & 0.3749 & 0.0415 & 0.0343 & 0.0234 & 0.0992 \\
\hspace*{3ex} \citet{cowan99}                & 0.158  & 0.146  & 0.135  & 0.439  & \nodata& \nodata& \nodata& \nodata \\
\hspace*{3ex} \citet{kratz04} (Fe-seed)      & 0.163  & 0.151  & 0.138  & 0.452  & \nodata& \nodata& \nodata& \nodata \\
\hline
$r$-residuals:                               &        &        &        &        &        &        &        &        \\
\hspace*{3ex} \citet{cowan99}                & 0.240  & 0.254  & 0.158  & 0.652  & \nodata& \nodata& \nodata& \nodata \\
\hspace*{3ex} \citet{beer01}                 & 0.178  & 0.171  & 0.133  & 0.482  & \nodata& \nodata& \nodata& \nodata \\
\enddata
\tablecomments{The $r$-residuals are calculated as 
$N_{r, \odot} = N_{\odot} - N_{s}$.}
\end{deluxetable}

%% file: tab9.tex
\begin{deluxetable}{ccclcccclccc} 
\tablecaption{Ages Derived from Individual Chronometers
\label{agestab}}
\tablewidth{0pt}
\tablecolumns{12}
\tablehead{
\colhead{} &
\colhead{} &
\colhead{} &
\multicolumn{4}{c}{``standard'' stars with \eps{La/Eu}~$<+0.25$} &
\colhead{} &
\multicolumn{4}{c}{stars with actinide boost} \\
\cline{4-7} \cline{9-12}
\colhead{Chronometer} &
\colhead{P.R.\ at} &
\colhead{log(P.R.)} &
\colhead{Observed} &
\colhead{No.} &
\colhead{Age} &
\colhead{Age spread} &
\colhead{} &
\colhead{Observed} &
\colhead{No.} &
\colhead{Age} &
\colhead{Age spread} \\
\colhead{pair} &
\colhead{time ``zero''} &
\colhead{} &
\colhead{mean (dex)} &
\colhead{stars} &
\colhead{(Gyr)} &
\colhead{(Gyr)} &
\colhead{} &
\colhead{mean (dex)} &
\colhead{stars} &
\colhead{(Gyr)} &
\colhead{(Gyr)} }
\startdata
Th/La & 0.585  & $-$0.233 & $-$0.67$\pm$0.03 ($\sigma$=0.10) & 16 & 20.4$\pm$4.1 & 4.7 & & $-$0.37$\pm$0.03 ($\sigma$=0.05) & 4 &    6.4$\pm$4.3 & 2.3 \\
Th/Eu & 0.463  & $-$0.334 & $-$0.56$\pm$0.03 ($\sigma$=0.08) & 16 & 10.6$\pm$4.1 & 3.7 & & $-$0.24$\pm$0.03 ($\sigma$=0.06) & 4 & $-$4.4$\pm$4.3 & 2.8 \\
Th/Er & 0.236  & $-$0.627 & $-$0.91$\pm$0.04 ($\sigma$=0.11) & 12 & 13.2$\pm$4.3 & 5.1 & & $-$0.66$\pm$0.08 ($\sigma$=0.17) & 4 &    1.5$\pm$5.5 & 7.9 \\
Th/Hf & 0.648  & $-$0.188 & $-$0.61$\pm$0.05 ($\sigma$=0.13) & 8  & 19.7$\pm$4.5 & 6.1 & & $-$0.26$\pm$0.06                 & 1 &    3.4$\pm$5.7 & \nodata \\
Th/Ir & 0.0677 & $-$1.169 & $-$1.42$\pm$0.06 ($\sigma$=0.15) & 8  & 11.7$\pm$4.8 & 7.0 & & $-$1.12$\pm$0.16 ($\sigma$=0.18) & 2 & $-$2.3$\pm$8.8 & 8.4 \\
Th/Pb & 0.111  & $-$0.955 & $-$1.21$\pm$0.23 ($\sigma$=0.32) & 2  & $>$9.9       & 11. & & $-$0.43$\pm$0.25                 & 1 & \nodata        & \nodata \\
\enddata
\tablecomments{No age is derived from the Th/Pb chronometer for the stars with an actinide boost 
because the decay of all actinide material present could never produce the observed Th/Pb ratio
in \cshill.}
\end{deluxetable}